\documentclass[twocolumn,twocolappendix]{aastex631}
\usepackage{CJK}

\usepackage{amsmath}
\usepackage{appendix}
\usepackage{enumitem}

\DeclareGraphicsExtensions{.eps,.eps.gz,.epsi}
\newcommand{\teff}{\mbox{$T_{\rm eff}$}}
\newcommand{\logg}{{\rm{log}~$g$}}

\newcommand{\feh}{{\rm [Fe/H]}} 
\newcommand{\ebv}{$E(B-V)$}

\shortauthors{Huang et al.}

\graphicspath{{./}{my_figures/}}

\begin{document}
\begin{CJK*}{UTF8}{gbsn}
\title{Stellar Loci. VIII. Photometric Metallicities for 100 Million Stars Based on Synthetic Gaia Colors}
\shorttitle{Photometric Metallicities for 100 Million Stars}
\author[0000-0002-1259-0517]{Bowen Huang (黄博闻)}
\affiliation{Institute for Frontiers in Astronomy and Astrophysics, Beijing Normal University, Beijing, 102206, China}
\affiliation{School of Physics and Astronomy, Beijing Normal University No.19, Xinjiekouwai St, Haidian District, Beijing, 100875, China}

\author[0000-0003-2471-2363]{Haibo Yuan (苑海波)}
\affiliation{Institute for Frontiers in Astronomy and Astrophysics, Beijing Normal University, Beijing, 102206, China}

\affiliation{School of Physics and Astronomy, Beijing Normal University No.19, Xinjiekouwai St, Haidian District, Beijing, 100875, China}

\correspondingauthor{Yuan Haibo (苑海波)}
\email{yuanhb@bnu.edu.cn}

\author[0000-0003-3535-504X]{Shuai Xu (徐帅)}
\affiliation{Institute for Frontiers in Astronomy and Astrophysics, Beijing Normal University, Beijing, 102206, China}
\affiliation{School of Physics and Astronomy, Beijing Normal University No.19, Xinjiekouwai St, Haidian District, Beijing, 100875, China}
\author[0000-0001-8424-1079]{Kai Xiao (肖凯)}
\affiliation{School of Astronomy and Space Science, University of Chinese Academy of Sciences, Beijing, 100049, China}
\author[0000-0002-5818-8769]{Maosheng Xiang (向茂盛)}
\affiliation{National Astronomical Observatories, Chinese Academy of Sciences, 20A Datun Road, Chaoyang District, Beijing, 100101, China}
\affiliation{Institute for Frontiers in Astronomy and Astrophysics, Beijing Normal University, Beijing, 102206, China}
\author[0000-0003-3250-2876]{Yang Huang (黄样)}
\affiliation{School of Astronomy and Space Science, University of Chinese Academy of Sciences, Beijing, 100049, China}
\author[0000-0003-4573-6233]{Timothy C. Beers}
\affiliation{Department of Physics and Astronomy, University of Notre Dame, Notre Dame, IN 46556, USA}
\affiliation{Joint Institute for Nuclear Astrophysics -- Center for the Evolution of the Elements (JINA-CEE), IN 46556, USA}

\begin{abstract} 
We apply the stellar locus method to synthetic $(BP-RP)_{XPSP}$ and $(BP-G)_{XPSP}$ colors derived from corrected Gaia BP/RP (XP) spectra to obtain precise estimates of metallicity for about 100 million stars in the Milky Way (34 million giants in the color range $0.6 < (BP-RP)_0 < 1.75$ and 65 million dwarfs in the color range  $0.2 < (BP-RP)_0 < 1.5$). 
The sub milli-magnitude precision of the derived synthetic stellar colors enables estimates of metallicity for stars as low as [Fe/H] $\sim -4$. Multiple validation tests indicate that the typical metallicity precision is between 0.05 -- 0.1 dex for both dwarfs and giants at [Fe/H] = 0 as faint as G $\sim$ 16, and decreases to 0.15 -- 0.25 dex at [Fe/H] = $-$2.0.  For $-4.0 <\feh < -3.0$, the typical metallicity precision decreases to on the order of 0.4 -- 0.5 dex, based on the results from the reference sample.
Our achieved precision is comparable to or better than previous efforts using the entire XP spectra, and about three times better than our previous work based on Gaia EDR3 colors.
This opens up new opportunities for investigations of stellar populations, the formation and chemical evolution of the Milky Way, the chemistry of stars and star clusters, and the identification of candidate stars for subsequent high-resolution spectroscopic follow-up. The catalog is publicly available at \href{url}{https://doi.org/10.12149/101548}.

\end{abstract}


\section{Introduction} 

A large, complete, and ideally unbiased census of stellar metallicities plays a crucial role in understanding the formation and evolution of stars and the Milky Way. In the past decades, stellar metallicities have been measured primarily from large-scale spectroscopic surveys. 
Notable surveys include the Sloan Digital Sky Survey and the Sloan Extension for Galactic Understanding and Evolution (SDSS/SEGUE; \citealt{SDSS2000}; \citealt{SEGUE2009}; \citealt{SEGUE2022}), the Radial Velocity Experiment (RAVE; \citealt{RAVE2006}), the Large Sky Area Multi-Object Fiber Spectroscopic Telescope (LAMOST; \citealt{LAMOST2012}; \citealt{LAMOST2014}), the Galactic Archaeology with HERMES project (GALAH; \citealt{GALAH2015}), the Apache Point Observatory Galactic Evolution Experiment (APOGEE; \citealt{APOGEE2017}), and the Gaia DR3 General Stellar Parametriser-spectroscopy (Gaia GSP-Spec; \citealt{GaiaGSP2023}) based on the Gaia Radial Velocity Spectrometer (RVS; \citealt{RVS2004} and \citealt{RVS2018}) spectra. Spectra with moderate to high resolution can yield precise metallicity estimates, typically around or surpassing 0.1 dex, particularly when the spectral signal-to-noise ratios are high. 
However, due to the time-consuming and complex data analysis that is required compared to photometric surveys, the cumulative number of observed sources from all the aforementioned surveys is limited to approximately 10 million stars. This represents only a tiny fraction of the estimated total number of stars in the Milky Way, equivalent to about one per myriad. The combination of under-sampling and intricate target-selection strategies presents challenges for obtaining a comprehensive understanding of the formation and evolution of the Milky Way.

Fortunately, the total number of sources provided by photometric surveys far exceeds that of spectroscopic surveys. Given that stellar colors are influenced not only by effective temperature, but also by other stellar atmospheric parameters such as metallicity (and to a lesser extent surface gravity), stellar metallicity can be estimated using photometric data. The metallicity dependence of a given color is influenced by the metallic absorption lines present within the transmission curve of its filters. This dependency is more pronounced for bluer colors, due to the prevalence of metallic absorption lines in the blue wavelength range. Similarly, the metallicity dependency is heightened for narrow-band filters that cover strong metallic absorption lines. Therefore, a series of previous efforts have utilized blue or narrow-band filters in the estimation of stellar metallicity.

Illustrative examples are the series of studies conducted using the SDSS or re-calibrated SDSS photometric data (\citealt{Yuan2015c}) in the $ugriz$-bands, as documented by \cite{SDSS_Metal2008}, \cite{Yuan2015a} (Paper I), \cite{Yuan2015b} (Paper III), and \cite{ZhangRY2021} (Paper IV). The typical precision of $\feh$ estimates in the aforementioned studies can reach 0.1 dex.\footnote{For completeness, we note that, in a series of papers, An et al. (\citealt{An2020, An2021a, An2021b, An2023}) have improved the identification and isolation of individual stellar populations in the Galactic halo through photometric metallicity estimates based on an updated set of empirically calibrated stellar isochrones in the SDSS and Pan-STARRS 1 systems.}

Likewise, \cite{HY2019,HY2022,HY2023} employed a similar technique to analyze photometric data from the SkyMapper Southern Survey DR1 and DR2 (SMSS; \citealt{SMSS2018DR1}, \citealt{SMSS2019DR2}, and \citealt{SMSScali}), as well as from the Stellar Abundances and Galactic Evolution Survey DR1 (SAGES DR1; \citealt{SAGES2023}). The optimally designed narrow-/medium-band $u,v$ filters utilized in these surveys facilitated  stellar metallicity estimates down to $\feh \sim - 3.5 $ for a combined total of approximately 50 million stars.  \cite{LX2024} (Paper VII）further employed the NUV-band of the Galaxy Evolution Explorer (GALEX; \citealt{GALEX2015}) to attain a typical  precision of approximately 0.11 dex for dwarfs and 0.17 dex for giants, encompassing an effective metallicity range as low as $\feh$ = $-$3.0 for dwarfs and $\feh$ = $-$4.0 for giants. Based on narrow-band photometry targeting the CaII HK lines, \cite{Pristine2017} and \cite{Pristine2023} have estimated stellar metallicities with a statistical precision of 0.11--0.14 dex
(primarily based on stars with $\feh > -1$); the available metallicity range of their sample is down to $\feh \lesssim$ $-$3.5.

The above efforts concentrated on the use of narrow- and medium-band metallicity-sensitive filters. In contrast, broad-band photometry is generally deemed less suitable for metallicity estimation due to its weak sensitivity on metallicity. However, based on very high-quality $G$, $BP$, and $RP$ broad-band photometric data from Gaia EDR3 (\citealt{GaiaEDR32021}), along with careful reddening corrections and color calibrations, \cite{XuShuai2022a} (Paper V) achieved a typical metallicity precision of $\sim$ 0.2 dex. This demonstrates that broad-band photometry remains capable of precise estimation of stellar metallicity, provided that the photometric precision is sufficiently high.

Gaia Data Release 3 (Gaia DR3; \citealt{GaiaDR3content}) has provided an extensive and innovative dataset, comprising approximately 220 million flux-calibrated low-resolution BP/RP (XP; hereafter) spectra covering the wavelength range of $336-1020$\,nm at a spectral resolving power of $R \sim 20-70$. Using the Gaia XP spectra, several all-sky catalogs of stellar metallicity have recently been published (\citealt{AndraeXPMetal}, \citealt{Pristine2023}, \citealt{ZhangXPMetal}, \citealt{FallowsXPMetal}, and \citealt{LJDXPMetal}). 

The Gaia XP spectra can also provide precise synthetic photometry (\citealt{GaiaXPsynColor}), making it a valuable dataset for photometric-metallicity estimation.
Here we demonstrate that synthesized Gaia DR3 $G$, $BP$, and $RP$ magnitudes (using corrected Gaia XP spectra (\citealt{HuangBW2024}), coupled with the stellar locus method (\citealt{Yuan2015a}) in the color-color diagram, demonstrates remarkably high precision, surpassing that of its Gaia DR3 photometry-based 
counterparts. Our study not only opens new possibilities for research in Galactic formation and evolution, but also sheds light on the optimal utilization of Gaia XP spectra for stellar-metallicity estimation.

This paper is organized as follows.
In Section\,2, we introduce the datasets used in our work.
In Section\,3, we describe the methods and results.
In Section\,4, we incorporate corrections on extinction, color, brightness, and the Hertzsprung-Russell Diagram (HRD).  In Section\,5, we compare our results with stellar metallicities from the literature, and further test our results using star clusters. In Section\,6, we present the final catalog and its statistical properties. A summary is provided in Section\,7.

\section{DATASETS} 
\subsection{Gaia DR3 and Corrected XP Spectra} 
The ESA Gaia mission is primarily dedicated to obtaining astrometry, photometry, and spectroscopy of objects within the Milky Way and the Local Group (\citealt{GaiaMission2016a}).
Gaia DR3 contains over 1.8 billion sources with photometry and 1.47 billion sources with at least five-parameter astrometry (two positions, the parallax, and two proper motion components;  \citealt{GaiaDR3content}).
It also provides approximately 220 million XP spectra covering the wavelength range $336-1020$\,nm (\citealt{GaiaDR3content}; \citealt{GaiaXPVad}).
Gaia XP spectra take the form of a projection onto 55 orthonormal Hermite functions as base functions for both the BP and RP spectra, rather than the flux as a function of wavelength (\citealt{dr3intcali}; \citealt{dr3extcali}). 

However, Gaia XP spectra have been found to suffer systematic errors that depend on the normalized spectral energy distribution and $G$ magnitude (\citealt{dr3extcali}; \citealt{HuangBW2024}). 
\cite{HuangBW2024} offer a comprehensive correction for the systematic errors in the wavelength space, within the ranges of approximately $-0.5<BP-RP<2$, $3<G<17.5$, and $\ebv <0.8$. Their tests demonstrate that they effectively rectified the systematic errors in the Gaia XP spectra, achieving an internal precision of 1--2 per cent.  Here we utilize the \texttt{GaiaXPy} package (\citealt{GaiaXPy}) to convert the coefficients into wavelength space, and then apply the aforementioned correction.

\subsection{LAMOST DR7} 
LAMOST, operated by the National Astronomical Observatory of China, Chinese Academy of Sciences, is a 4-meter quasi-meridian reflective Schmidt telescope equipped with 4000 fibers. The LAMOST Data Release 7 (DR7) \footnote{http://dr7.lamost.org/v1.3/} contains a total of 10,431,197 low-resolution spectra spanning the entire optical wavelength range of $369-910$\,nm, with a spectral resoving power of $R \sim 1800$.
It also provides approximately 6.1 million stellar atmospheric parameters, such as effective temperature ($\teff$), surface gravity (\logg), and metallicity ($\feh$) for 4.6 million unique sources. 
These fundamental parameters are derived 
by the LAMOST Stellar Parameter Pipeline (\citealt{LASP};\citealt{LAMOST2015}).
Internal errors in the LAMOST parameters, for spectra with signal-to-noise ratios exceeding 20, are typically $\teff ~\sim 50-100$\,K, \logg $~\sim 0.05-0.1$\,{dex} and $\feh~\sim 0.05-0.1$\,dex \citep{Niu2021a}. The colors and magnitudes of these sources range roughly over $0<BP-RP<3$ and $9<G<17.8$. \cite{Niu2023} have calibrated the internal systematic errors of stellar chemical abundances for FGK-type dwarf stars using wide binaries (for more details, refer to their Figure 4). The underestimation of LAMOST \feh\ can be as high as 0.4 dex at 4000\,K and 0.1 dex at 7000\,K. We employ this calibration to our reference sample of dwarfs.

\subsection{PASTEL}
The PASTEL catalog (\citealt{PASTEL2010,PASTEL2016}) is a comprehensive bibliographical compilation of stellar atmospheric parameters, including (\teff, \logg, and \feh), derived from the analysis of high-resolution, high signal-to-noise spectra using model atmospheres. As of February 2016, the PASTEL catalog encompassed 31,401 distinct stars, accounting for 64,082 records. For stars with multiple measurements, this work adopts the average values. According to the standard deviation of multiple measurements for a single star (\citealt{BeyoundSpectroscopy1}), the precision is better than 0.05 dex for stars with $\feh > -2.0$, and ranges between 0.05 and 0.10 dex for stars with $\feh \leq -2.0$.

\section{Methods and Results} 
\subsection{Initial Data Selection}
We first use the corrected Gaia XP spectra 
and the transmission curves of Gaia DR3 ($G$, $BP$ and $RP$) to obtain synthetic magnitudes $G_{XPSP}$, $BP_{XPSP}$, and $RP_{XPSP}$.
Subsequently, we cross-match LAMOST DR7 and PASTEL with Gaia DR3 XP spectra, with a matching radius of 1 arcsec. A total of 3.8 million common sources from LAMOST and 15,328 common sources from PASTEL are found. Duplicate sources within LAMOST are identified, and only the observation with the highest signal-to-noise ratio is retained.

Reddening correction is performed using the dust reddening map of \cite{SFD98}, hereafter SFD98, and  reddening- and temperature-dependent extinction coefficients from \cite{ZhangRYExtinction2023}. 
To optimize the use of the SFD98 map and ensure the data quality, we employ the following initial criteria to all the common sources:
\vskip 2 cm
\begin{enumerate}[itemsep=0.2cm, topsep=0.5cm, parsep=0.2cm, partopsep=0.3cm]
    \item Absolute values of Galactic latitude are greater than 20\degr.
    \item Vertical distances to the Galactic disk are greater than 0.2 kpc. Here the geometric distances from \cite{BailerJonesDist} are employed.
    \item $\ebv _{ \rm SFD98 } < 0.5$. 
    \item The signal-to-noise ratios in the $g$-band ($SNR_g$) of LAMOST spectra are greater than 20 for LAMOST sources.
    \item $phot\_bp\_rp\_excess\_factor <0.02 \times (BP_{XPSP}-RP_{XPSP})^2+0.055 \times (BP_{XPSP}-RP_{XPSP})+1.165$. The $phot\_bp\_rp\_excess\_factor$ is defined as $C = \frac{I_{BP}+I_{RP}}{I_{G}}$ to evaluate the background and
contamination issues in both $BP$ and $RP$ photometry and spectra. Our empirical cut helps to exclude sources with low-quality Gaia measurements.
\end{enumerate} 

\begin{figure*}[ht]
\includegraphics[width=180mm]{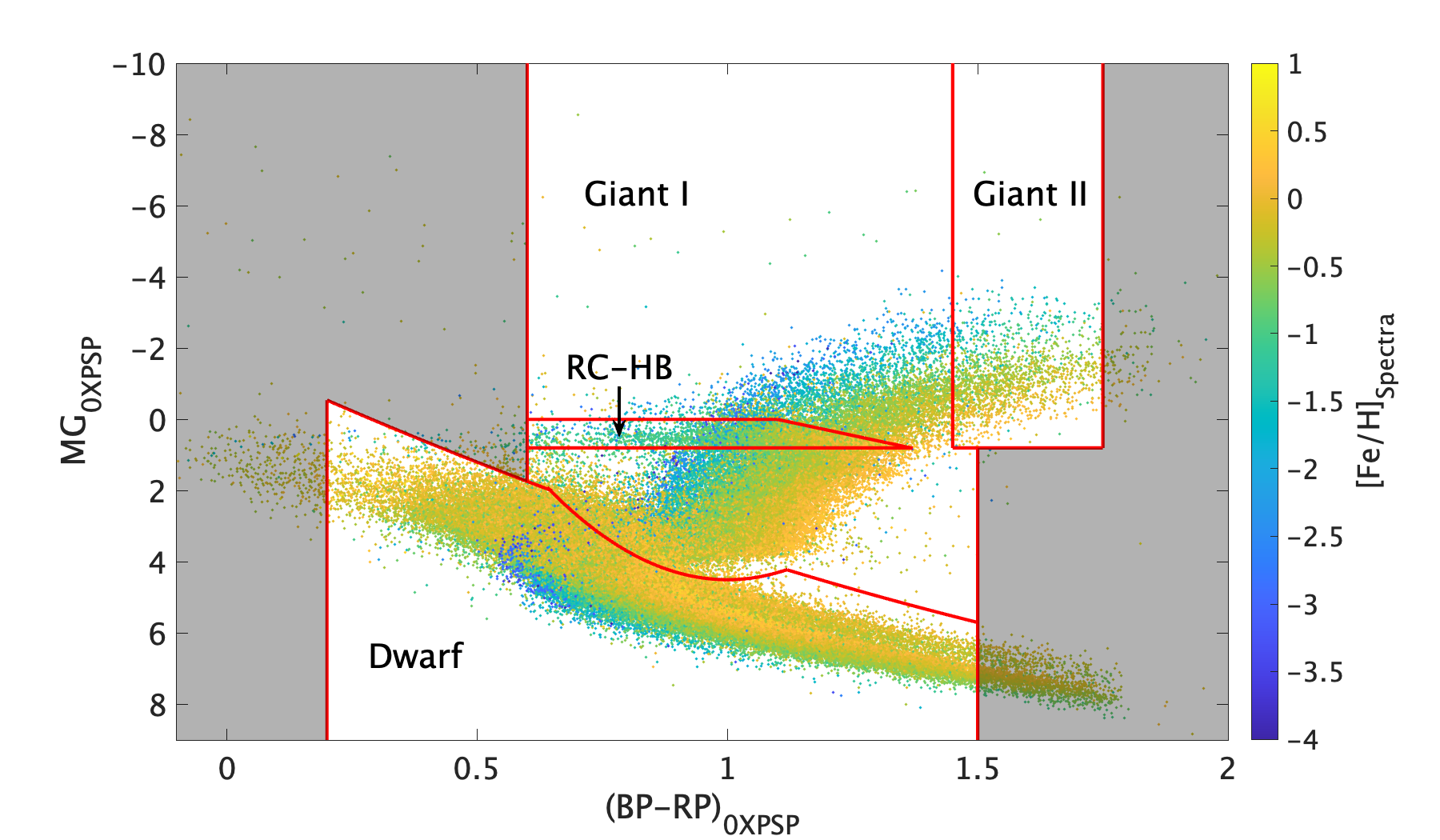}
\caption{Hertzsprung–-Russell Diagram (HRD) of the initially selected sample.
The colors of the points indicate the \feh~ from the included sources. The colors of the points reflect their metallicity, as coded on the color bar on the right.  The HRD is partitioned into four distinct regions by solid red lines: Dwarf, RC-HB, Giant\,I, and Giant\,II (see text). 
} 
\label{Fig_partition}
\end{figure*}

Following application of the above criteria, 1.4 million LAMOST sources and 1,548 PASTEL sources remain. 
The HRD of the selected sample is presented in Figure~\ref{Fig_partition}.
We partition this sample into four sets within the HRD: dwarf stars, red clump and horizontal-branch (RC-HB) stars，Giant I stars, and Giant II stars. The division criteria are:

\begin{gather}
    Dwarf: \notag \\ 
    0.2 \leq (BP-RP)_{0XPSP} \leq 1.5 \notag \\
    MG_{0XPSP} > -(BP-RP)_{0XPSP}^2 \notag \\ 
    +6.5 \times (BP-RP)_{0XPSP}-1.8 \notag \\
    MG_{0XPSP}>-20 \times (BP-RP)_{0XPSP}^2 \notag \\
    +40 \times (BP-RP)_{0XPSP}-15.5 \notag \\
    \notag \\ 
    RC-HB: \notag \\ 
    (BP-RP)_{0XPSP}>0.6 \notag \\
    0.0<MG_{0XPSP}<0.8 \notag \\
    MG_{0XPSP}> 3 \times (BP-RP)_{0XPSP}-3.3 \notag \\
    \notag \\ 
    Giant\,I: \notag \\ 
    0.6\leq(BP-RP)_{0XPSP}\leq1.45 \notag \\
    \text{not in the Dwarf and RC-HB regions} \notag \\
    \notag \\ 
    Giant\,II: \notag \\ 
    1.45<(BP-RP)_{0XPSP}\leq 1.75 \notag\\
    MG_{0XPSP} \leq 0.8 \notag\\
\label{4regions}
\end{gather}

Here $(BP-RP)_{0XPSP}$ and $(BP-G)_{0XPSP}$ represent the de-reddened synthetic colors of $BP-RP$ and $BP-G$, respectively. Similarly, $MG_{0XPSP}$ denotes the de-reddened absolute magnitude in the $G$-band.

\subsection{Reference Sample Selection}
Similar to the methodology employed in \cite{XuShuai2022a}, we further incorporate $12<G_{XPSP}<14$ and $\ebv < 0.03$ to select the reference sample of the best photometric quality and lowest reddening correction. Due to the limited number of PASTEL sources,  only $\ebv<0.1$ is required for the PASTEL sources. 
Furthermore, considering that LAMOST \feh~ is only valid down to $\sim -2.5$ and there is a 0.2\,dex systematic \feh~ difference between the LAMOST and PASTEL sources when $\feh < -0.75$, we exclude LAMOST sources with $\feh < -0.75$ to ensure the robustness of the stellar-loci fitting technique in the very metal poor (VMP; $\feh \leq$ $-2.0$), and extremely metal poor (EMP; $\feh \leq $ $-3.0$) regions. 
When $\feh \ge -0.75$, the overall offset between LAMOST and PASTEL converges to zero. The limited number of common sources between LAMOST and PASTEL makes it difficult to perform more precise corrections during the merging of the reference samples, and as a result, no such corrections have been implemented. Given that the number of sources from LAMOST is significantly greater than that from PASTEL in the $\feh \ge -0.75$ range, the LAMOST sample predominantly determines the behavior of the stellar loci at higher metallicities.

With these cuts, the reference sample consists of 148,419 sources in the Dwarf set, 4,816 sources in the RC-HB set, 40,723 sources in the Giant\,I set, and 366 sources in the Giant\,II set.

\subsection{Stellar Loci Fitting in the Dwarf, Giant I, and RC-HB Regions}
Similar to \cite{XuShuai2022a}, 
we fit $(BP-G)_{0XPSP}$ as a function of $(BP-RP)_{0XPSP}$ and $\feh$ using polynomials based on the least-squares method for the Dwarf, Giant I, and RC-HB sets, respectively. 
As demonstrated in Equations (2--5), where X represents $(BP-RP)_{0XPSP}$, Y represents $\feh$, Z represents $MG_{0XPSP}$, and $k = -5.25$, we decompose the fit into three components.
The $f_1$ represents the \feh-dependent variations of $(BP-G)_{0XPSP}$, $f_2$ accounts for the $(BP-G)_{0XPSP}$ variations influenced by $(BP-RP)_{0XPSP}$, and $f_3$ describes the $MG_{0XPSP}$-dependent variations, which is only used for the Giant\,I region. 

Note that the color of a star exhibits weaker sensitivity to metallicity when the star is more metal poor, eventually making it impossible to distinguish between different metallicities at the metal-poor end with a given photometric precision. To resolve this issue, we introduce a constant $k = -5.25$ into $f_1$, ensuring that the sensitivity diminishes progressively at the metal-poor end and eventually reaches zero at $\feh = -5.25$. 
The $f_2$ variable specifically refers to the stellar locus at $\feh = 0$. Due to the wide range of color in the Dwarf region, it is very challenging to achieve millimagnitude-level fitting using low-order polynomials. Therefore, the $f_2$ term for the Dwarf region is presented numerically in Appendix A rather than as polynomial coefficients.

In accordance with the findings reported in \cite{LX2024}, a moderate correlation has been identified between the fitting residual and the absolute magnitude in the Giant\,I region. Consequently, we incorporate $MG_{0XPSP}$ into the fitting process for the Giant\,I set; no correlation is observed in the Dwarf and RC-HB sets. 
Note that we only consider low-order terms of $MG_{0XPSP}$ to enhance our results, and the primary information for \feh~ comes from $(BP-RP)_{0XPSP}$ and $(BP-G)_{0XPSP}$ colors. For comparison, we also provide a fit model for the Giant\,I set excluding $MG_{0XPSP}$. The results are similar, as can be seen in the following outcomes.

\begin{table*}
\centering
\caption{Stellar Loci Fitting Coefficients}
\label{Tab1} 
\begin{tabular}{*{7}{r}} \hline\hline
Region & $a_1$ & $a_2$ &$a_3$ & $a_4$ & \\ \hline
Dwarf & $-$0.110 & $+$0.207 & $-$0.082 & $+$0.036 & \\
Giant I  & $-$0.032 & $-$0.047 & $+$0.183 & $-$0.052 & \\
Giant I* & $-$0.007 & $-$0.114 & $+$0.233 & $-$0.060 & \\
RC-HB & $-$0.209 & $+$0.472 & $-$0.275 & $+$0.067 & \\
Giant II & $+$0.580 & $-$2.662 & $+$3.996 & $-$1.936 & \\
\hline
\end{tabular}

\begin{tabular}{*{8}{r}} \hline\hline
Region & $b_1$ & $b_2$ &$b_3$ & $b_4$ & $b_5$ & $b_6$ & \\ \hline
Dwarf &\dots &\dots &\dots &\dots &\dots &\dots & \\
Giant I  & $-$0.016 & $-$0.502 & $-$3.085 & $+$97.723 & $+$333.558 & $-$9.799 & \\
Giant I* & $+$0.567 & $-$7.937 & $+$29.734 & $+$41.596 & $+$371.842 & $-$17.850 & \\
RC-HB & $+$429.437 & $-$1977.437 & $+$3539.140 & $-$2984.593 & $+$1635.689 & $-$223.656 & \\
Giant II & $+$0.622$ \times 10^3$ & $-$5.553$ \times 10^3$ & $+$15.910$ \times 10^3$ & $-$22.711$ \times 10^3$ & $+$16.789$ \times 10^3$ & $-$6.717$ \times 10^3$ & \\

\hline
\end{tabular}

\begin{tabular}{*{6}{r}} \hline\hline
Region & $c_1$ & $c_2$ &  $c_3$ & $c_4$ & \\ \hline
Dwarf &0 &0 &0 &0 & \\
Giant I &0 &0 &0 &0 & \\
Giant I*  & $-$0.777 & $+$0.404 & $+$0.252 & $+$0.031 & \\
RC-HB &0 &0 &0 &0 & \\
Giant II & $+$1.317$ \times 10^4$ & $-$1.314$ \times 10^4$ & $+$0.593$ \times 10^4$ & $-$0.495$ \times 10^4$ & \\
\hline
\end{tabular}

Giant I* indicates the fitting model with  $MG_{0XPSP}$ included.
\end{table*}

\begin{gather}
  (BP-G)_{0XPSP} = (f_1+f_2+f_3)/1000 \\
    f_1 = 
    \begin{bmatrix}
        a_1 \\
        a_2 \\
        a_3 \\
        a_4 \\
    \end{bmatrix}
    ^T 
    \begin{bmatrix}
        X ^3 \\
        X ^2  \\
        X \\
        1 \\
    \end{bmatrix}
    (
    \begin{bmatrix}
        Y ^3 \\
        Y ^2  \\
        Y \\
    \end{bmatrix}
    ^T
    \begin{bmatrix}
        1 \\
        -3k  \\
        3k^2 \\
    \end{bmatrix}
    ) \\
    f_2 = 
    \begin{bmatrix}
        b_1 \\
        b_2 \\
        b_3 \\
        b_4 \\
        b_5 \\
        b_6 \\
    \end{bmatrix}
    ^T 
    \begin{bmatrix}
        X ^5 \\
        X ^4  \\
        X ^3  \\
        X ^2  \\
        X   \\
        1 \\
    \end{bmatrix}
    \\
    f_3 = 
    \begin{bmatrix}
        c_1 \\
        c_2 \\
        c_3 \\
        c_4\\
    \end{bmatrix}
    ^T
    \begin{bmatrix}
        Z \\
        X Z \\
        X ^2 Z  \\
        Z ^2 \\
    \end{bmatrix}
\label{FitModel_func_2}
\end{gather}

\begin{figure*}[ht]
\includegraphics[width=180mm]{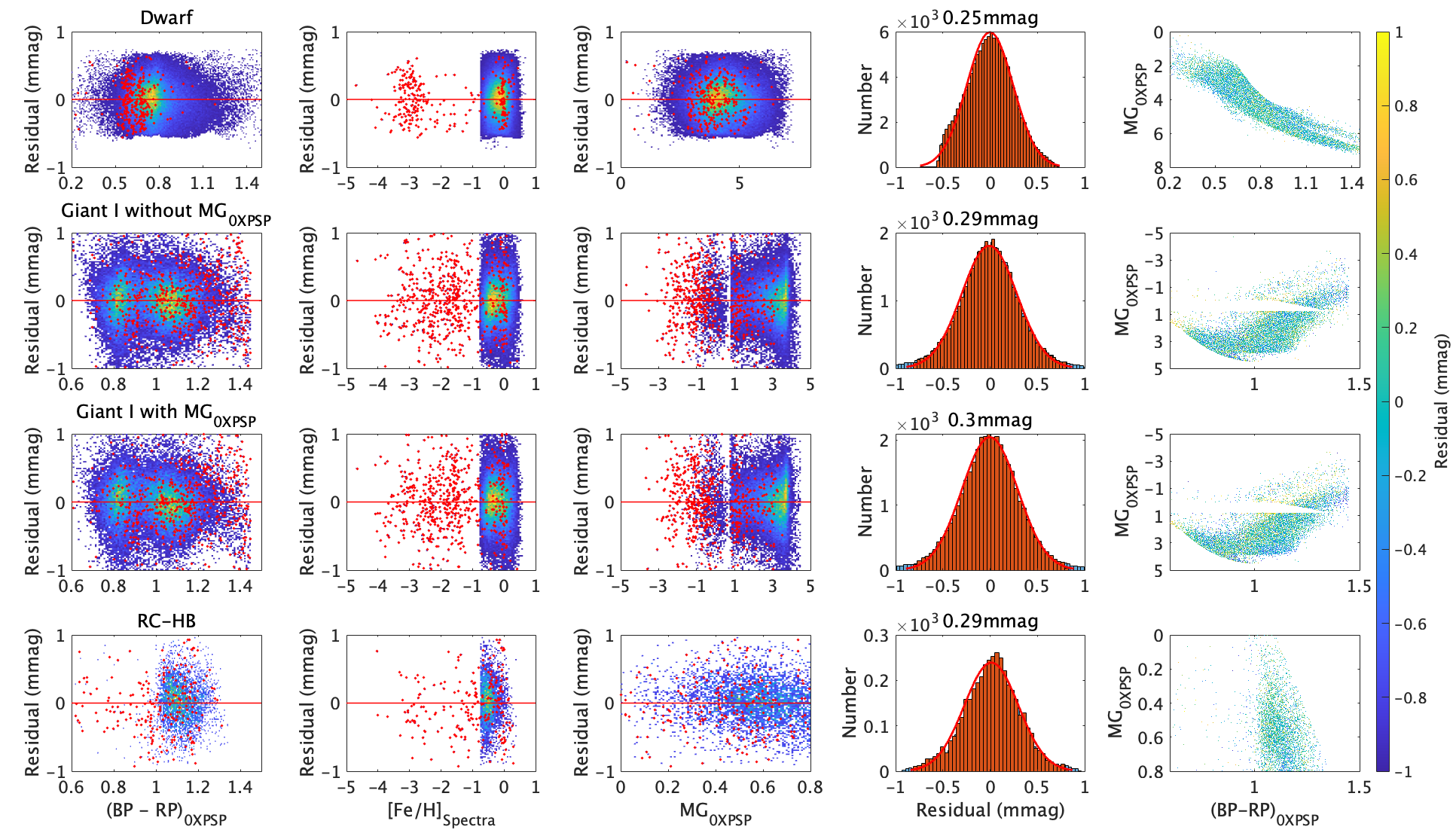}
\caption{Stellar loci fitting residuals. From top to bottom are shown the Dwarf region，the Giant\,I region without and with $MG_{0XPSP}$, and the RC-HB region, respectively. The first, second, and third columns show the residuals, as functions of $(BP_{XPSP}-RP_{XPSP})_0$, $\feh _{Spectra}$, and $MG_{0XPSP}$, respectively, with the red points indicating PASTEL sources.
The fourth column illustrates the distributions of residuals, with the red bar employed to compute the standard deviation of the Gaussian distribution, mitigating the impact of a limited number of outliers.
The fifth column displays the distribution of residuals in the HRD, where the color of the points corresponds to the residuals, as coded on the color bar to the right.} 
\label{FitResult}
\end{figure*}

\begin{figure*}[ht]
\includegraphics[width=180mm]{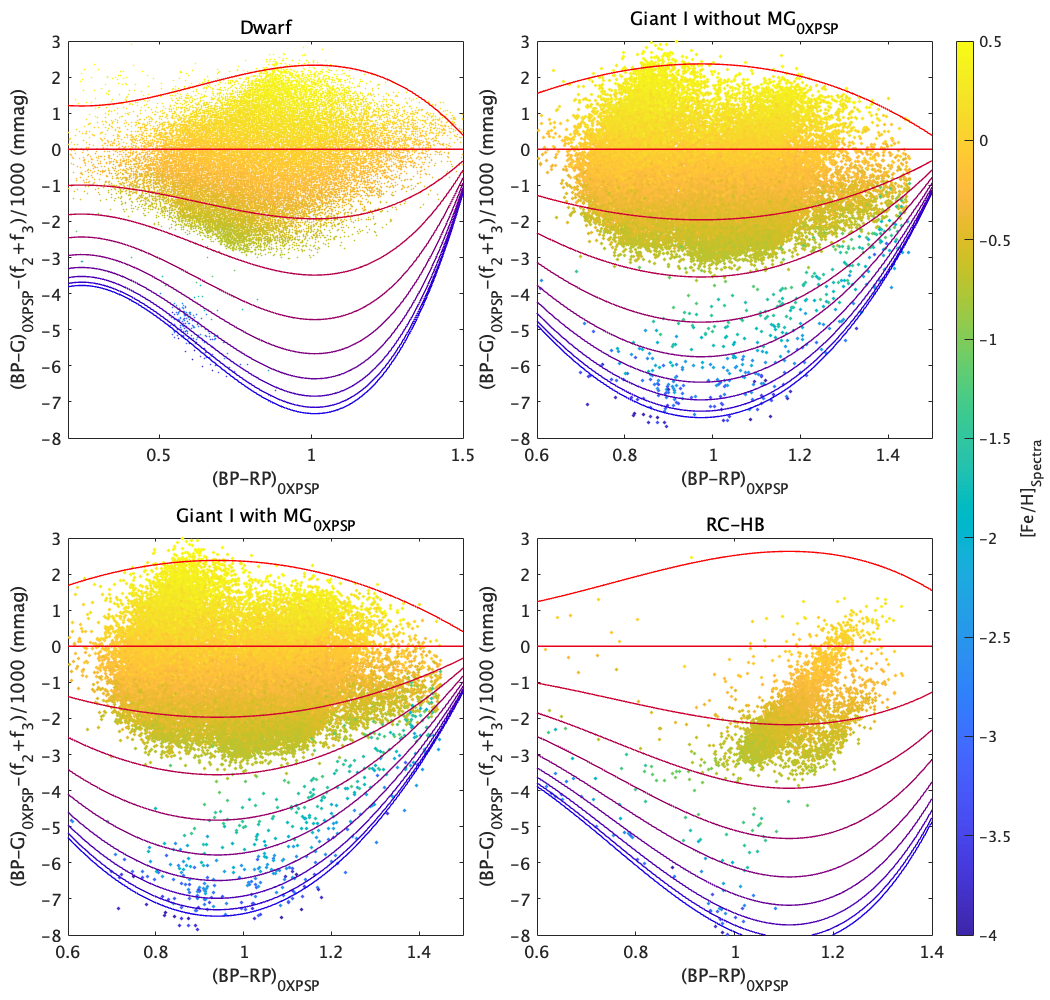}
\caption{Distributions of the reference sample in $(BP_{XPSP}-RP_{XPSP})_0$ vs. $(BP_{XPSP}-G_{XPSP})_0 - (f_2+f_3)/1000$, color-coded by \feh, as shown by the color bar to the right. The lines from top to bottom represent our best fits of $f_1/1000$ for \feh\ values of $+0.5, 0.0, -0.5, -1.0, -1.5, -2.0, -2.5, -3.0, -3.5$ and $-4.0$, respectively.} 
\label{FitModel}
\end{figure*}

The fitting process employs multiple iterations of $3\,\sigma$ clipping to eliminate erroneous sources. Additionally, in the Dwarf region,  $2\,\sigma$ clipping is applied to eliminate binary stars. The fitting coefficients are presented in Table~\ref{Tab1}.  The fitting residuals are depicted in 
Figure~\ref{FitResult}, showing extremely small dispersion values of 0.25 --  0.30 mmag for all sets. No observable correlation with $(BP-RP)_{0XPSP}$, \feh, and $MG_{0XPSP}$ is found.
Figure~\ref{FitModel} shows the color-color diagram depicting $(BP-RP)_{0XPSP}$ versus $(BP-G)_{0XPSP} - (f_2+f_3)/1000$. It focuses on the $(BP-G)_{0XPSP}$ variations that solely depends on \feh.
As shown in Figure~\ref{FitModel}, our fitting model ensures the monotonicity of \feh~ with respect to $(BP-G)_{0XPSP}$ in the range of $\feh > -5.25$. 
The stellar loci exhibit very weak sensitivity to \feh~ when $(BP-G)_{0XPSP} \sim1.5$. Given the obvious correlation between \feh~ and $MG_{0XPSP}$ illustrated in Figure~\ref{Fig_partition}, we incorporate $MG_{0XPSP}$ in the subsequent section to derive \feh~in the Giant\,II region.

\begin{figure*}[ht]
\includegraphics[width=180mm]{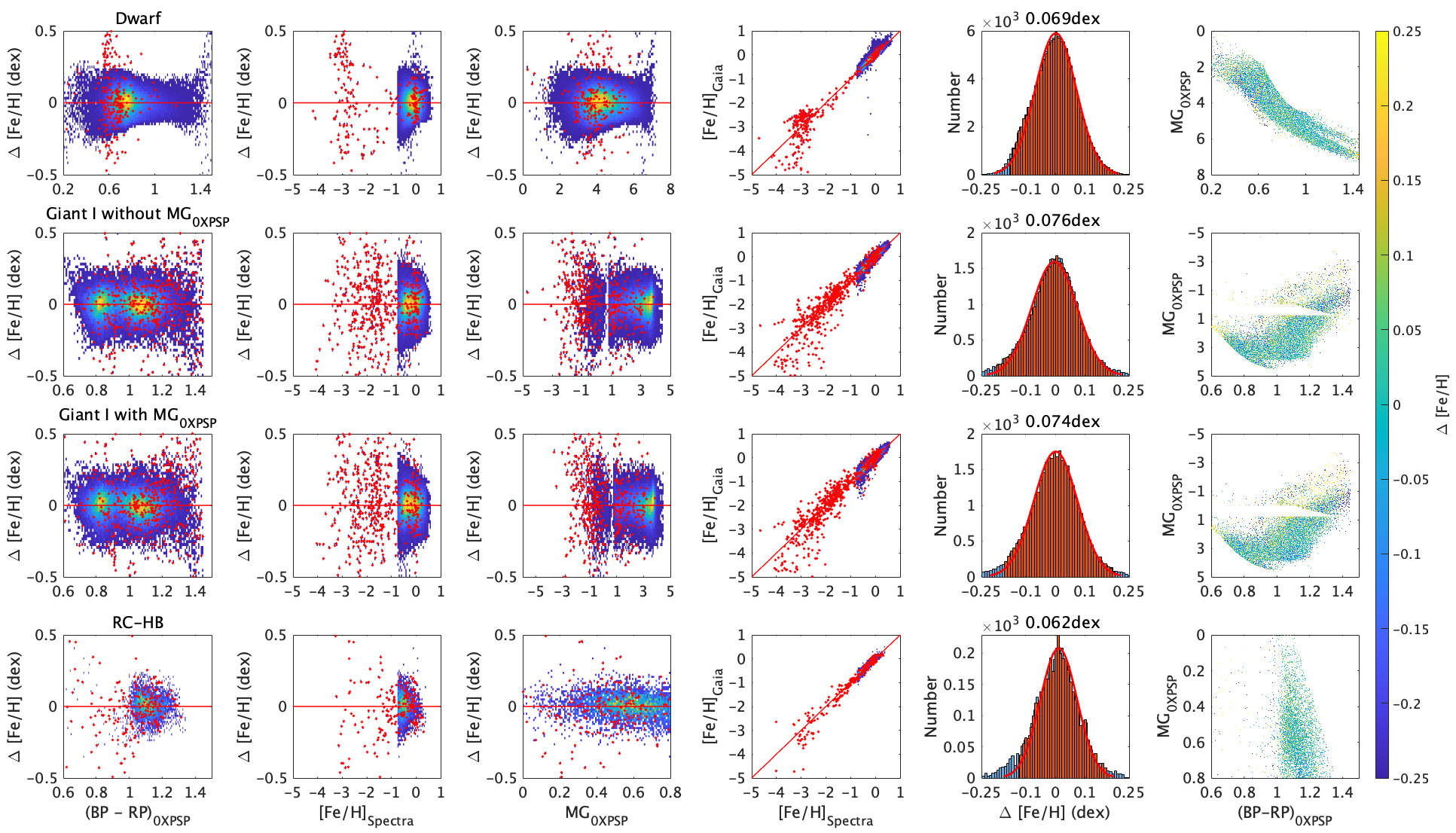}
\caption{Similar to Figure~\ref{FitResult}, but for the comparisons between the $\feh_{\rm Gaia}$ and 
$\feh_{\rm Spectra}$ of the reference sample, where the color of the points corresponds to the residuals, as coded on the color bar to the right. } 
\label{FehResult}
\end{figure*}

By integrating our polynomial model with $(BP-RP)_{0XPSP}$, $(BP-G)_{0XPSP}$, and $MG_{0XPSP}$, we further derive the estimated [Fe/H]$_{\text{Gaia}}$ by solving Equations (2)--(5).
Figure~\ref{FehResult} shows the comparison between [Fe/H]$_\text{Gaia}$ and [Fe/H]$_\text{Spectra}$ for the reference sample. The dispersion of $\Delta \feh$ for each set in the reference sample is only 0.06--0.08\,dex; our result ensures consistency and robustness with \feh~ extending into the EMP region ($-4 < \feh\ < -3$).

\subsection{Stellar Loci Fitting in the Giant\,II Region}

\begin{figure*}[ht]
\includegraphics[width=180mm]{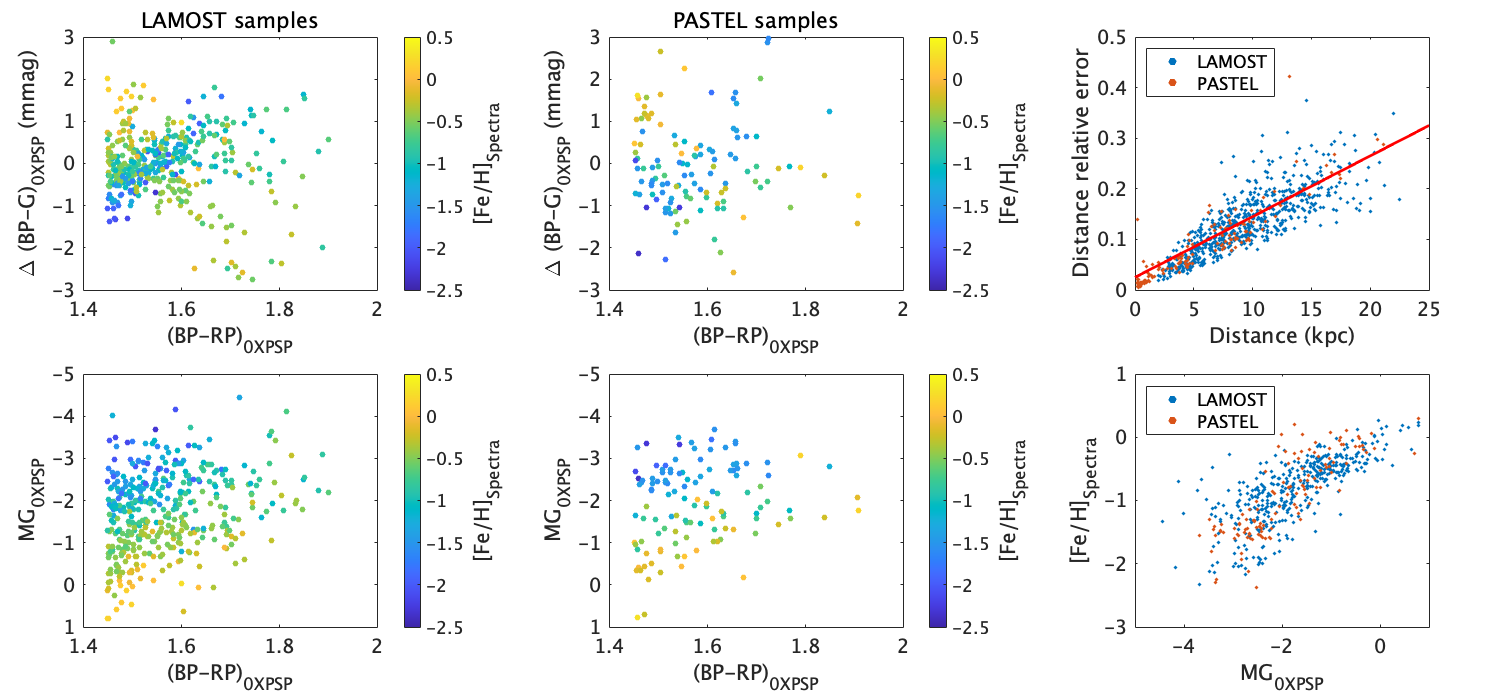}
\caption{The color-color diagrams of $(BP-RP)_{0XPSP}$ versus $(BP-G)_{0XPSP}$, along with the HRDs of sources in the Giant II region. 
The left column represents the sources from LAMOST, while the middle column represents the sources from PASTEL.
The colors of the points indicate the \feh~ values, coded by the color bars.  The term $\Delta\,$ $(BP-G)_{0XPSP}$ represents the residuals derived from the quadratic function fitting of $(BP-RP)_{0XPSP}$ to $(BP-G)_{0XPSP}$, aiming to characterize the sensitivity of $(BP-G)_{0XPSP}$ to \feh. 
The upper right panel illustrates the relationship between the relative distance errors and distance, with the red solid line employed to exclude sources with significant relative distance errors. The lower right panel depicts the relationship between \feh\ and $MG_{0XPSP}$ after excluding these sources.} 
\label{Giant_II_property}
\end{figure*}

For the reasons mentioned in subsection 3.3, we treat the Giant\,II Region separately. As illustrated in Figure~\ref{Giant_II_property}, the distinction in \feh~ is not apparent in the color-color diagram of $(BP-RP)_{0XPSP}$ and $(BP-G)_{0XPSP}$. However, in the HRD, $MG_{0XPSP}$ exhibits noticeable sensitivity to \feh. Thus, in this subsection, we primarily use the information from $MG_{0XPSP}$ to estimate \feh.  

Due to the limited number of reference sources in the Giant\,II region, we merge sources from both the LAMOST and PASTEL datasets, and refrain from applying the $\feh \ge -0.75$ criterion to the LAMOST sample.
We also extend the color range of the reference sample to $1.45 < (BP-RP)_{0XPSP} \leq 2$ to ensure that the polynomial fit remains stable at the reddest edge of the Giant II region.
We further eliminate sources with large distance errors from \cite{BailerJonesDist} by 
$\frac{r\_hi\_geo - r\_lo\_geo}{2 \times r\_med\_geo} < r\_med\_geo \times 0.000012+0.025$
(the red solid line in the top right panel of Figure~\ref{Giant_II_property}) to ensure the accuracy of $MG_{0XPSP}$.

\begin{figure*}[ht]
\includegraphics[width=180mm]{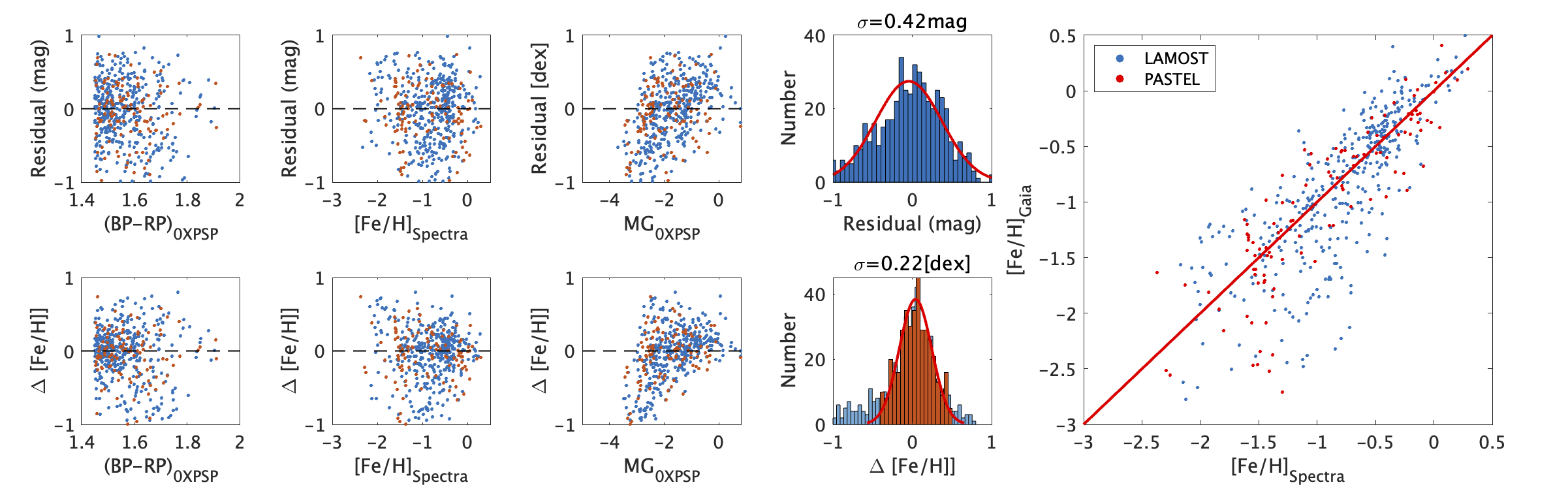}
\caption{Fitting residuals in the Giant\,II region. } 
\label{Giant_II_fit}
\end{figure*}

The fitting model and process in the Giant II region are similar to those in other regions. However, we utilize $(BP-RP)_{0XPSP}$, $(BP-G)_{0XPSP}$, and $\feh$ to fit $MG_{0XPSP}$. We let $MG_{0XPSP} = (f_1+f_2+f_3)$, where X represents $(BP-RP)_{0XPSP}$, Y represents $\feh$, Z represents $(BP-G)_{0XPSP}$ and $k = -5.25$.
Final fitting results are shown in Figure~\ref{Giant_II_fit}. The fitting residuals exhibit no discernible correlation with the independent variables, and maintain robustness at $\feh = -2 $.

\section{Corrections} 
\subsection{Corrections on Extinction, Color, and Brightness}
In the previous section, we employed samples with the highest precision in Gaia ($12<G<14$) and minimal extinction to model the mapping from $(BP-RP)_{0XPSP}$, $(BP-G)_{0XPSP}$, and $MG_{0XPSP}$ to \feh.
However, extrapolating these models to stars of wider brightness ranges and larger extinction might introduce systematic biases.
The extinction coefficients from \cite{ZhangRYExtinction2023} are primarily applicable within the \ebv~ range of 0--0.5 mag and the \teff~ range of 4,000--10,000\,K (approximately $0 < (BP-RP)_{0XPSP} < 1.61$). Furthermore, the extinction coefficients from \cite{ZhangRYExtinction2023} are based on Gaia DR3 photometry. The discrepancy between Gaia DR3 photometry and synthetic photometry introduces a small error in these coefficients when applied to this work, subsequently affecting the metallicity estimation.
Hence, we utilize samples filtered through the initial criteria (Section 3.1) to implement supplementary corrections to the models, encompassing a broader brightness ($3<G<17.65$) and extinction ($\ebv~< 0.5$) range.
For $\ebv~> 0.5 $, \cite{ZhangRYExtinction2023} adopted the extinction coefficients at $\ebv~= 0.5$. 
Consequently, sources of $\ebv~> 0.5 $ in our sample are subjected to relatively inaccurate extinction corrections. We similarly rectify this instance of high extinction ($\ebv~> 0.5$) using the aforementioned samples. However, the reliability of this correction remains unverified, and we acknowledge this caveat in the final catalog. 

To correct for the aforementioned systematic biases, we select samples from LAMOST and PASTEL, respectively. For the LAMOST sample, we apply the initial criteria (see Section 2.2), but exclude the $\ebv _{SFD} < 0.5$ restriction, and require $\feh >-0.75$. 
Due to the brightness limitations of the LAMOST sample, we use the PASTEL sample to supplement the brightness-related corrections when $G < 10$. For the PASTEL sample, we apply the first and fifth of the initial criteria, and additionally require $\ebv _{SFD} < 0.1$.

\begin{figure*}[ht]
\includegraphics[width=180mm]{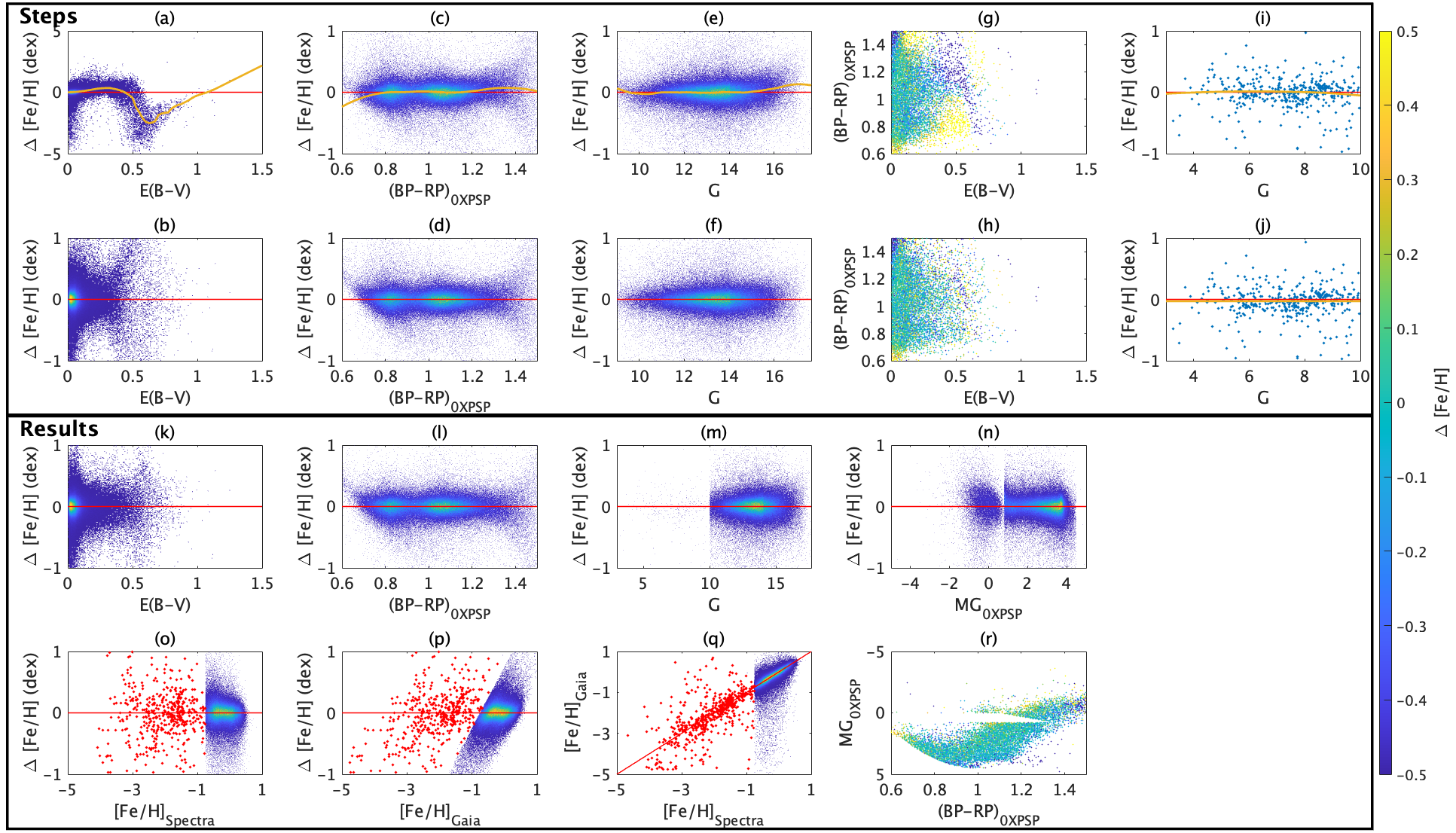}
\caption{The correction on extinction, colors, and brightness with $MG_{0XPSP}$ in the Giant\,I region. Panels (a) to (h) sequentially illustrate the correction of systematic biases based on extinction, color, brightness, and the simultaneous dependence of color and extinction using the LAMOST sample. Panels (i) and (j) show the correction for brightness-dependent systematic biases when $G < 10$ based on the PASTEL sample. The second row of panels present the results after each correction step.
Panels (k) to (r) show the final residuals on \ebv, $(BP-RP)_{0XPSP}$, $G$ , $MG_{0XPSP}$, $[Fe/]_{\rm{Spectra}}$, $\feh_{\rm{Gaia}}$, and HRD, respectively. The yellow solid lines in panels (a), (c), (e), and (i) represent the corrections. In panels (g), (h), and (r), the colors of the dots indicate $\Delta$ $\feh$, as coded on the color bar to the right. 
The red dots in panels (o), (p), and (q) are sources from the PASTEL sample, while the background dots are sources from the LAMOST sample.}
\label{Giant_correction_with_MG}
\end{figure*}

Figure~\ref{Giant_correction_with_MG} presents the correction procedures for the Giant I Region with $MG_{0XPSP}$ as an example. The correction steps and results for the other regions are similar. Panels (a) to (h) sequentially illustrate the correction of systematic biases on extinction, color, brightness, and the simultaneous dependence on color and extinction using the LAMOST sample stars. The second row of panels present the results after each correction step. 
Additionally, panels (i) and (j) show the correction for brightness-dependent systematic biases when $G < 10$ based on the PASTEL sample.
Panels (k) to (r) display the final results following all five correction steps. 

As shown in Figure~\ref{Giant_correction_with_MG}, the primary systematic biases stem from extinction, while the color and brightness-dependent systematic errors are minimal. Comparing panel (b) with panel (k) reveals that, after correcting for the simultaneous dependence on color and extinction, the large dispersion of $\Delta$\,\feh~ at \ebv~ $\sim 0.5$ is significantly reduced, underscoring the necessity of this correction. After applying the above corrections, we do not observe any other significant systematic biases. However, panel (r) reveals a small systematic bias structure in the HRD (a similar structure is also found in the RC-HB region) , which we discuss and address in the next subsection.

\subsection{Corrections in the HRD}
\begin{figure*}[ht]
\includegraphics[width=180mm]{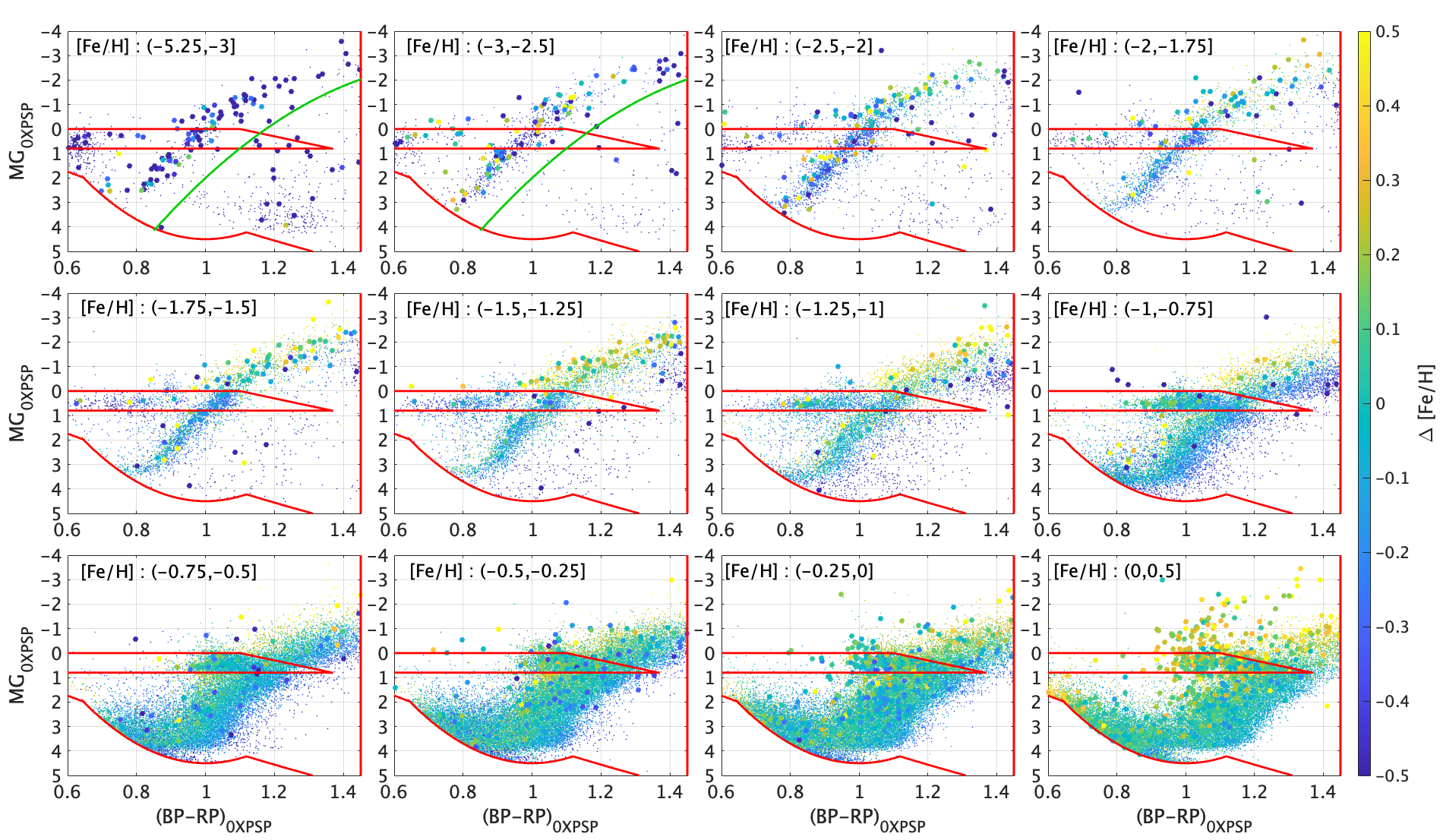}
\caption{The Hertzsprung-Russell Diagrams of the Giant I and RC-HB regions in different \feh~ranges. The colors of the dots indicate the difference in \feh~between this work and LAMOST (small dots) as well as PASTEL (large dots), as coded on the color bar to the right. The solid green lines in the upper left two panels distinguish the regions of red stragglers or subgiants.} 
\label{Giant_overall_HRD_sct}
\end{figure*}

In this subsection, we focus on the systematic bias structure of the Giant I and RC-HB regions in the HRD. To address this, we selected samples from LAMOST and PASTEL. For the LAMOST sample, we apply the initial criteria (see Section 2.2). For the PASTEL sample, we use the same sample as in the previous subsection. As shown in Figure~\ref{Giant_overall_HRD_sct}, we divide the sample into 12 different sub-samples based on our \feh~ results, and find that the systematic bias structure in the HRD varies across each sub-sample. For a given sub-sample, most sources are concentrated near the isochrone, and do not exhibit systematic bias. Therefore, our correction in the HRD primarily addresses systematic bias structures at other locations. 

For each sub-sample we create a grid in the HRD. For each grid point, we select sources from LAMOST (small dots) within the ranges $ \Delta (BP-RP)_{0XPSP} < 0.1 $ and $ \Delta (MG)_{0XPSP} < 0.4 $, and from PASTEL (large dots) within the ranges $ \Delta (BP-RP)_{0XPSP} < 0.25 $ and $ \Delta (MG)_{0XPSP} < 1 $. We then fit the systematic bias using a second-order function of $(BP-RP)_{0XPSP}$ and $(MG)_{0XPSP}$ to obtain corrections at the grid point locations. For certain grid points, where the number of LAMOST sources is fewer than 10, we extend the range for selecting LAMOST sources to match the larger range used for PASTEL. 

\begin{figure*}[ht]
\includegraphics[width=180mm]{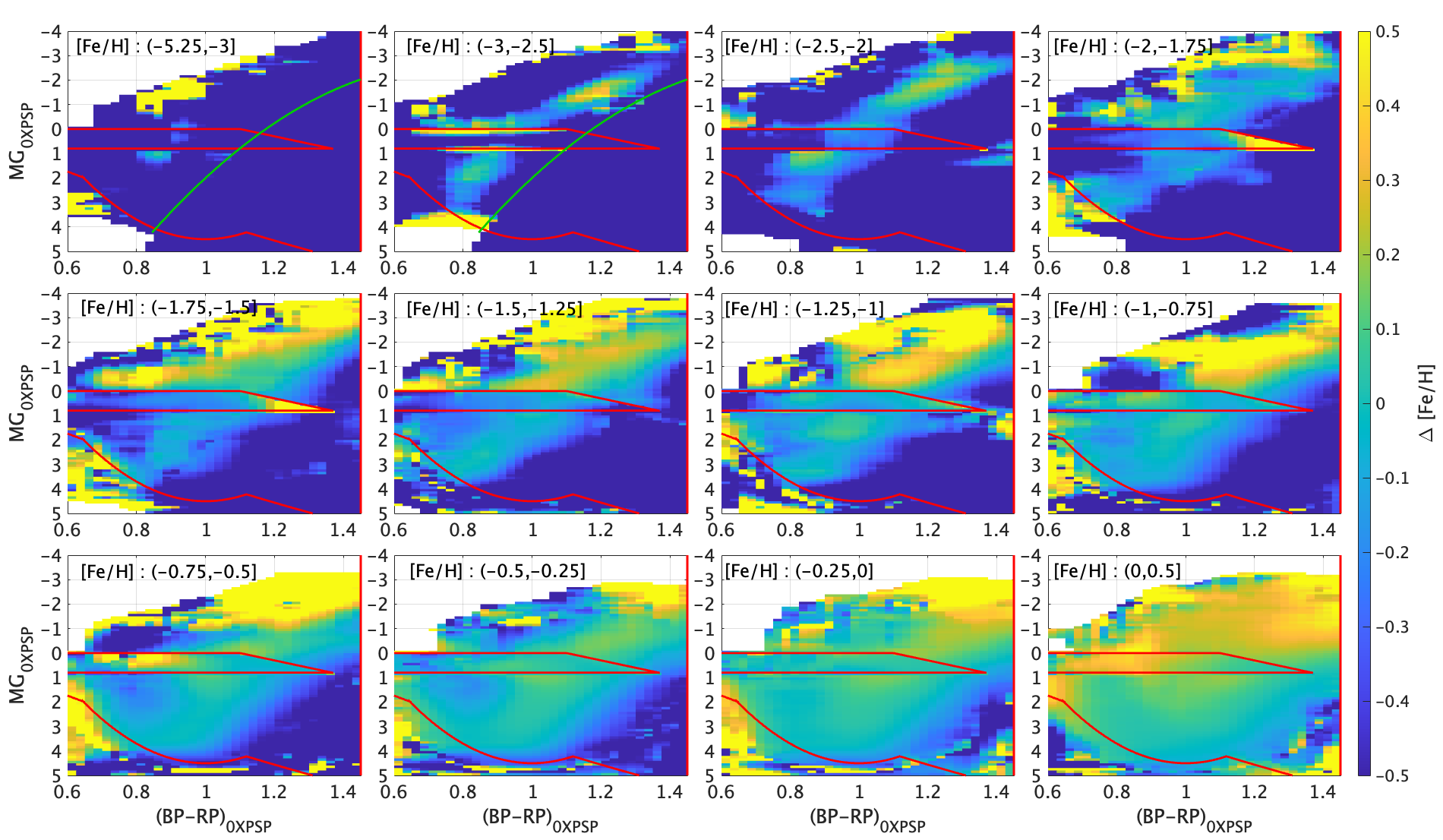}
\caption{Same as Figure~\ref{Giant_overall_HRD_sct}, but showing the result of the correction. The colors of the dots indicate the difference in \feh~between this work and LAMOST (small dots) as well as PASTEL (large dots), as coded on the color bar to the right.} 
\label{Giant_overall_HRD_grd}
\end{figure*}

Our correction result is shown in Figure~\ref{Giant_overall_HRD_grd}. For a source with a given $(BP-RP)_{0XPSP}$, $MG_{0XPSP}$, and \feh, we provide a correction using 3D interpolation. Note that for sources with $\feh < -2.75$, applying this correction to sources near the isochrone will cause large errors, due to the small number of sources used to generate the correction and the fact that LAMOST's \feh~ range only goes as low as $\sim -2.5$. Thus, in the range of $\feh < -2.75$, we only apply the correction to sources that are significantly off the isochrones (bottom right of the solid green lines). These sources are most likely red stragglers or subgiants, and not VMP or EMP stars. 

Based on the results of this subsection, we find that, even with the introduction of low-order polynomials in $MG_{0XPSP}$ in the stellar locus fitting, fine systematic bias structures are still observable (refer to the rightmost panels in Figure A3 of \cite{LX2024}).
We infer that the effect of absolute magnitude in stellar loci fitting (see the Appendix in \citealt{LX2024}) cannot be fully described by low-order polynomials.

\subsection{Identification of Binaries in the HRD for the Dwarf Region}
\begin{figure*}[ht]
\includegraphics[width=180mm]{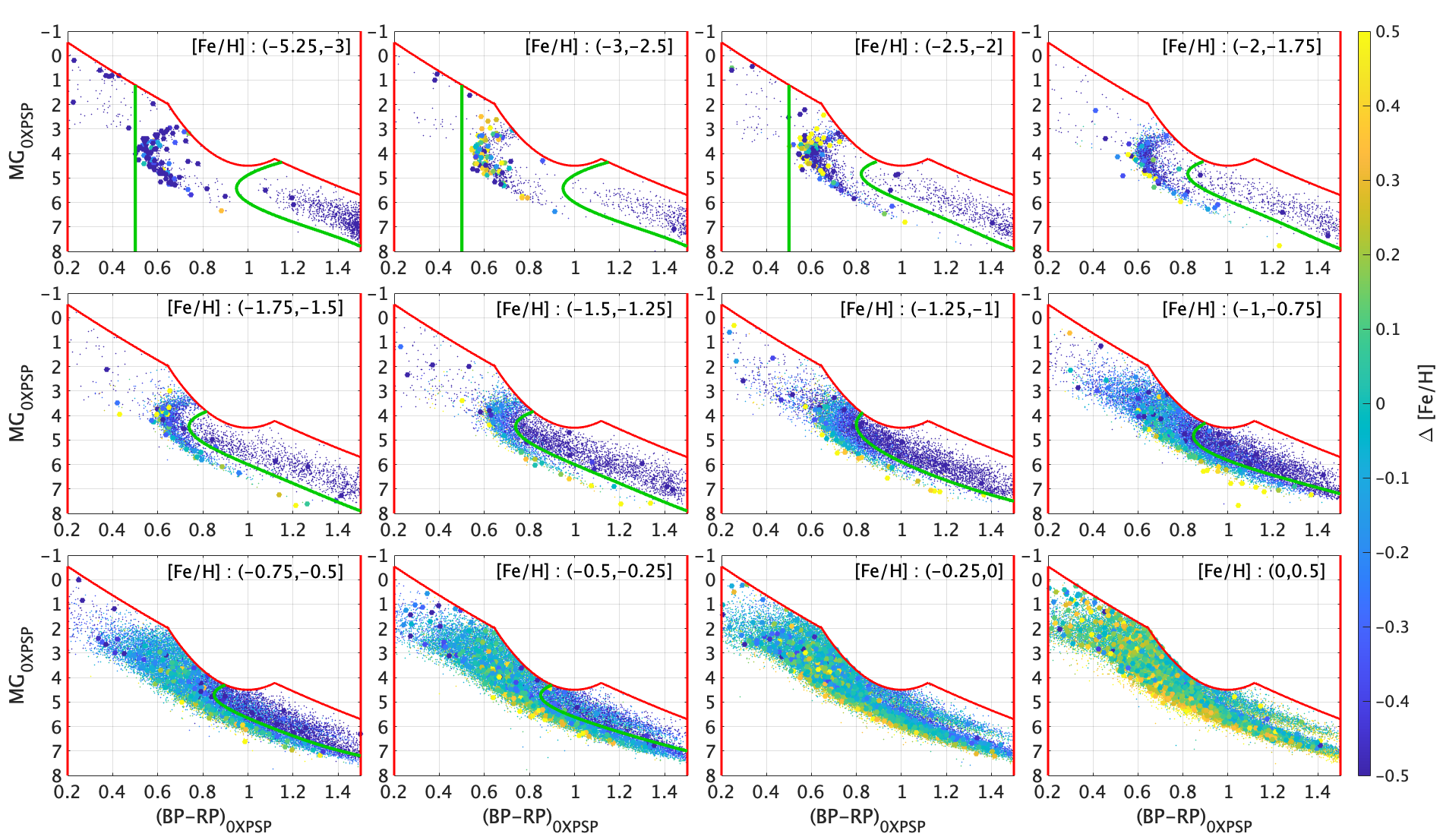}
\caption{Same as Figure~\ref{Giant_overall_HRD_sct}, but for the Dwarf region. The solid green curves distinguish the regions of likely main-sequence binaries. The green vertical solid lines distinguish the regions of likely blue stragglers. The colors of the dots indicate the difference in \feh~between this work and LAMOST (small dots) as well as PASTEL (large dots), as coded on the color bar to the right.} 
\label{Fig_Dwarf_Binary}
\end{figure*}

We use the method described in the previous subsection to examine the Dwarf region, aiming to identify main-sequence binaries. Main-sequence binaries are usually estimated to have lower metallicity using photometric data (see, e.g., Paper III; \citealt{Yuan2015b} and Paper V; \citealt{XuShuai2022a}).  Figure~\ref{Fig_Dwarf_Binary} shows that if sources with \feh~from our work are lower than \feh~from LAMOST and PASTEL, then their positions in the HRD are usually above the main sequence. This suggests that these sources are very likely binaries, as indicated to the right side of the empirical green curves in Figure~\ref{Fig_Dwarf_Binary}. 
Additionally, sources positioned to the left of the green vertical solid lines ($(BP-RP)_{0XPSP} < 0.5$, $\feh < -2$) are also considered to be influenced by binaries, such as blue stragglers. The final catalog will include a flag to preliminarily identify binaries within the dwarf region. Sources situated to the left of the green vertical solid lines will also be assigned the same flag to help mitigate contamination of VMP stars.

\vskip 1cm
\section{Comparisons and Tests} 
We employ the stellar loci fitting results and the above corrections to determine the photometric metallicity estimates, denoted as $\feh_{\rm Gaia}$.  In this section, we test and verify the accuracy and precision of our result.

\subsection{Comparison with SDSS and SEGUE}
We compare our results and those of \cite{AndraeXPMetal} and \cite{ZhangXPMetal} against the parameters from the SDSS Data Release 12 (DR12; \citealt{SDSS DR 12}).
Since the reference samples used in this work, as well as training samples from \cite{AndraeXPMetal} and \cite{ZhangXPMetal}, are not sourced from SDSS, this comparison serves as an independent test.

The SDSS DR12 represents the final data release of the SDSS-III, encompassing all SDSS observations until 2014 July.
Their parameter catalog comprises approximately 0.4 million sources with three fundamental stellar parameters ($\teff$, \logg,  and $\feh$). These parameters are obtained by the Sloan Extension for Galactic Understanding and Exploration (SEGUE) Stellar Parameter Pipeline (SSPP; \citealt{SEGUE pipeline}; \citealt{SEGUE Lee 2008a,SEGUE Lee 2008b,SEGUE Lee 2011,SEGUE Lee 2013}), spanning a temperature range of 4,000 to 10,000\,K. 
After applying the initial criteria and an additional requirement of $RUWE <1.1$ (see \citealt{Xu2022b}, Figure 5), there are 121,829 common sources between the four catalogs. 

We proceed to select sources with $G<17$, and categorize the common sources into giants and dwarfs based on the criteria delineated in Figure~\ref{Fig_partition}. In the case of dwarfs, we further adopt the approach described in the preceding section to eliminate main-sequence binaries.  

\begin{figure*}[ht]
\includegraphics[width=180mm]{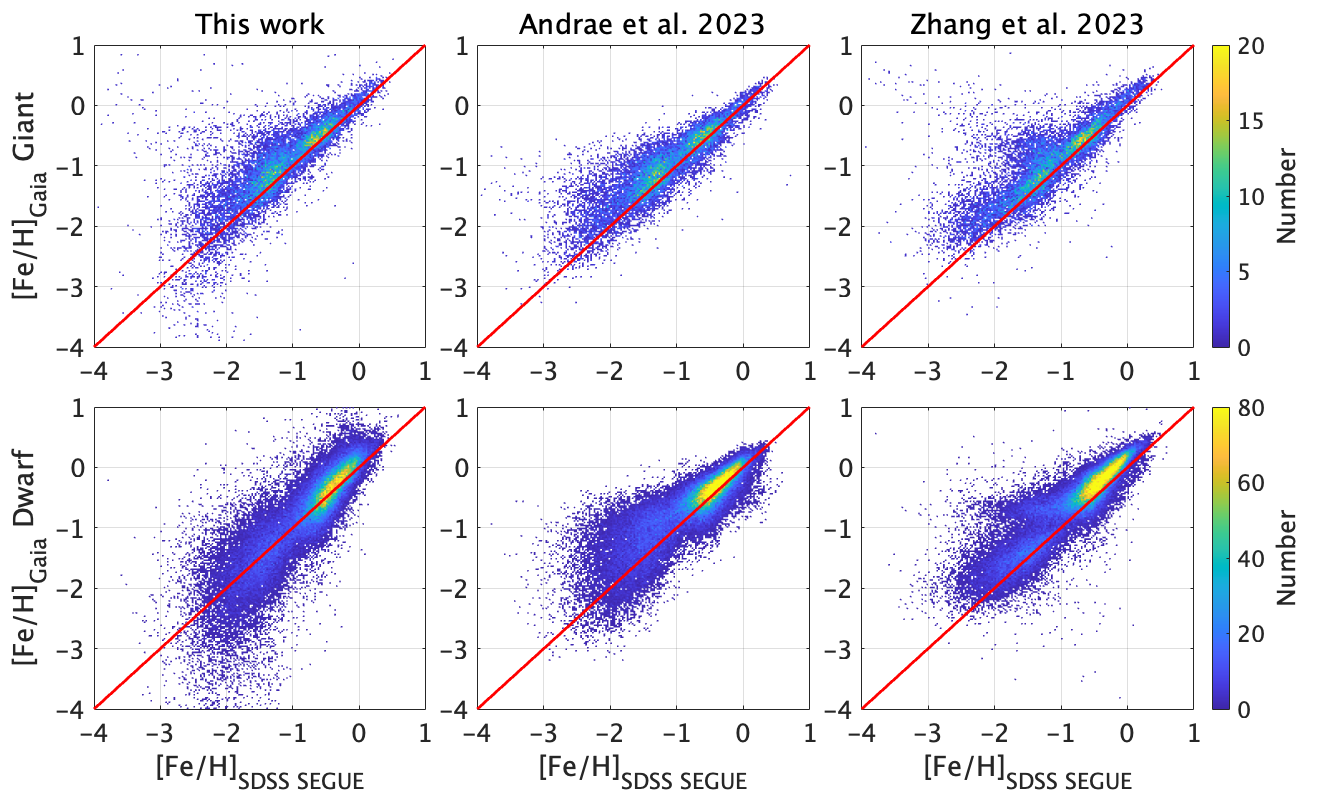}
\caption{Comparison of \feh~between this work (left panels), \cite{AndraeXPMetal} (middel panels), and \cite{ZhangXPMetal} (right panels) against SDSS DR12. The top and bottom panels
are for giants and dwarfs, respectively.  The colors represent the density of points.} 
\label{Compare_SDSS}
\end{figure*}

The comparison results are presented in Figure~\ref{Compare_SDSS}. Our results are in favorable agreement with SDSS, particularly for VMP stars.
This suggests that that VMP and EMP stars can be selected with high purity and completeness from our catalog. 
Furthermore, comparisons with metallicity estimates derived from high-resolution spectra, such as APOGEE DR17, GALAH DR4, and Gaia GSP-spec, are presented in Appendix B. These comparisons also demonstrate good agreement with the metallicity estimates from high-resolution, further supporting the above conclusion.

\subsection{The Influence of Carbon Enhancement}
Photometric-metallicity measurements are often overestimated for VMP/EMP stars, due to contamination from 
molecular bands of carbon and nitrogen (e.g., \citealt{Hong2024}, \citealt{HY2023}, and \citealt{LX2024}). 
In this subsection, we use the medium- and high-resolution spectroscopic catalog of VMP/EMP stars with available $\feh$ and $\rm [C/Fe]$ measurements compiled by \cite{Hong2024} to test whether our results are influenced by carbon enhancement. We also perform the same test on the results of \cite{AndraeXPMetal} and \cite{ZhangXPMetal}. 

\begin{figure*}[ht]
\includegraphics[width=180mm]{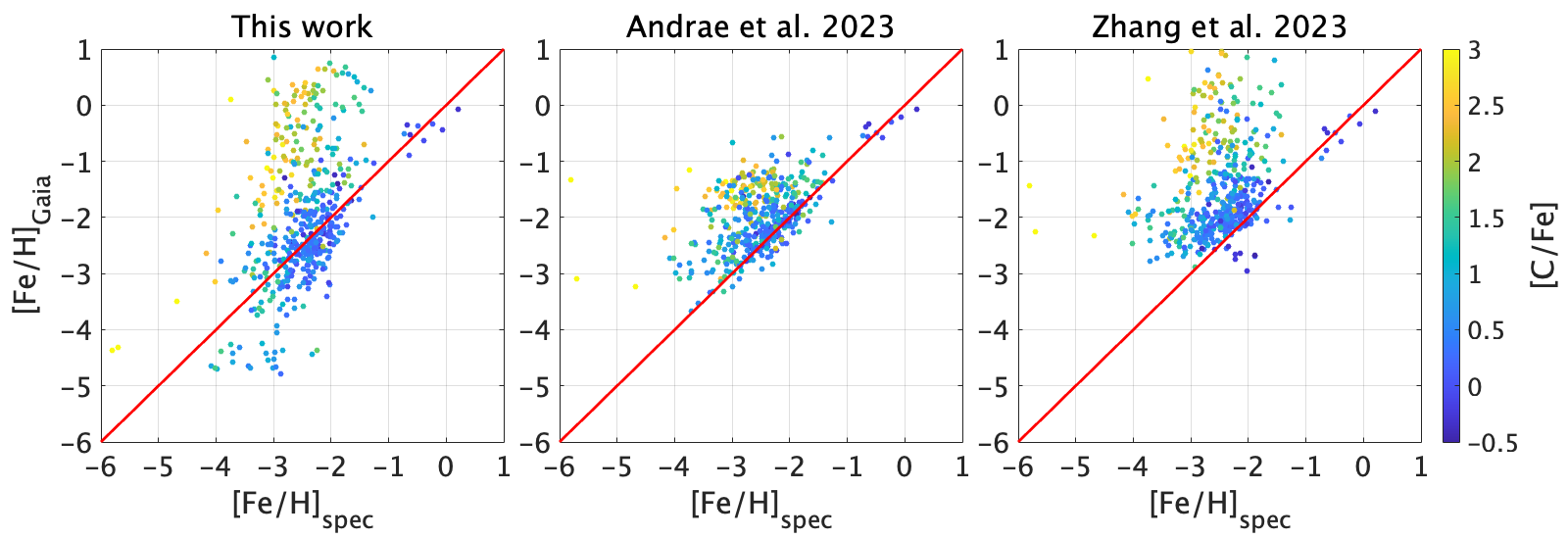}
\caption{Comparison of $\feh$ between the three Gaia XP spectra-based catalogs and the one compiled by \cite{Hong2024}. 
The colors of the dots denote $\rm [C/Fe]$, as coded on the color bar to the right.} 
\label{Compare_poor}
\end{figure*}

After application of the third and fifth of the initial criteria, there are 485 common sources. The comparison results are presented in Figure~\ref{Compare_poor}. All three catalogs are influenced by carbon enhancement, with the result of \cite{AndraeXPMetal} being relatively less affected. 
In addition, our result demonstrates better robustness for normal VMP stars, as already shown in the previous subsection.
It is noteworthy that our results tend to be systematically lower than those from [Fe/H]$_{spec}$ in the EMP region. One contributing factor is the non-Gaussian uncertainty associated with photometric metallicity estimation, particularly in the regime of extremely low sensitivity. Another factor arises from our assumption that metallicity sensitivity gradually diminishes at the metal-poor end, ultimately reaching zero at [Fe/H] = $-$5.25. While these assumptions enable robust discrimination in the extreme low-sensitivity regime, they may also introduce artificial systematic biases, which could be corrected in future studies.

\subsection{Comparison with LAMOST DR7}

In this subsection, we utilize LAMOST's $\feh$ as a reference to examine the precision of three Gaia XP spectra-based results with respect to brightness. 
We apply the initial criteria and further require that sources are not identified as binaries in section 4.3, and have $RUWE < 1.1$. There are 235,605 common sources for giants and 752,384 common sources for dwarfs.

\begin{figure*}[ht]
\includegraphics[width=180mm]{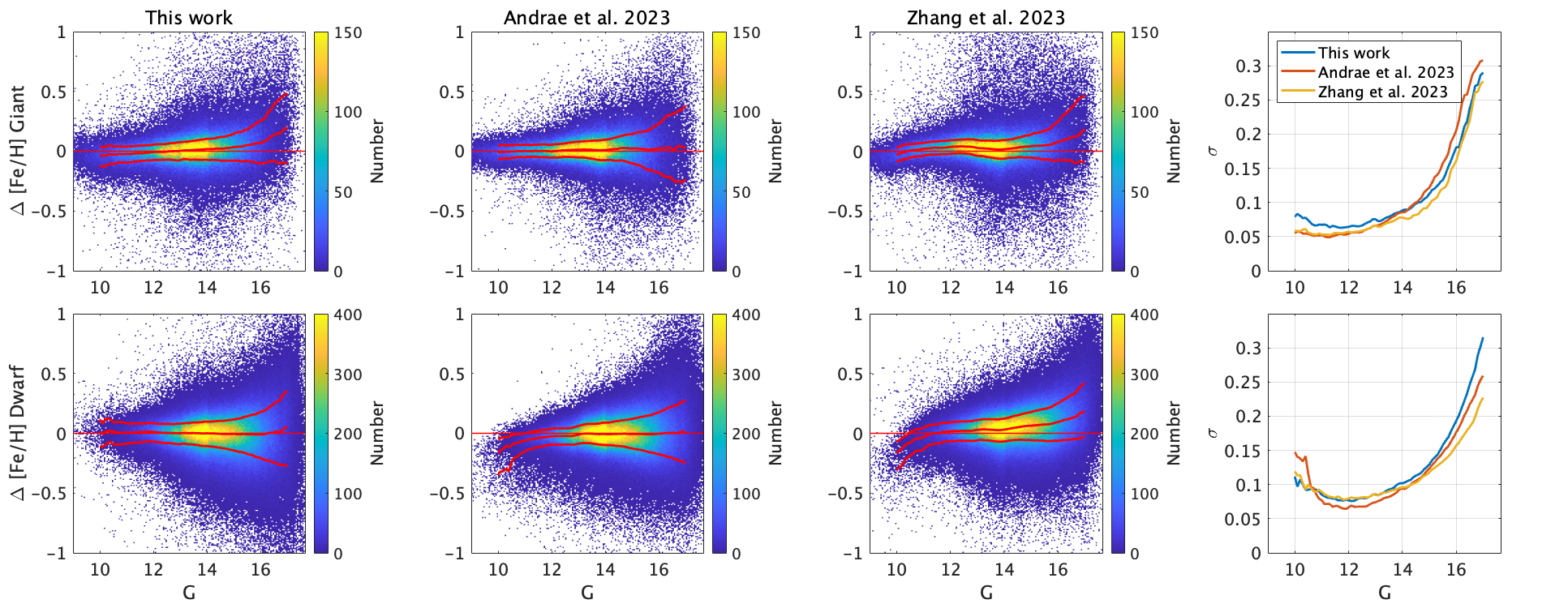}
\caption{The differences of $\feh$ estimates between the three Gaia XP spectra-based results and LAMOST DR7. The top and bottom panels are for giants and dwarfs, respectively.
In each panel, the red horizontal line indicates where $\Delta , \feh$ equals 0; the other three red lines indicate the 16th, 50th, and 84th percentiles, as functions of $G$. The colors represent the density of points.  
The rightmost column shows the dispersion of $\Delta \, \feh$, as a function of $G$. The dispersion of 
$\Delta \, \feh$ is calculated by taking half the difference between the 84th and 16th percentiles.} 
\label{Compare_LAMOST_G}
\end{figure*}

Figure~\ref{Compare_LAMOST_G} shows there are no significant systematic errors associated with the $G$ magnitude in our results, and the precision of our result is comparable to the other two Gaia XP spectra-based results. For giants, slight differences are observed between our results and those from LAMOST for faint sources.
For dwarfs, the comparison of our results and the dispersion of $\Delta \, \feh$ can be significantly affected by binaries. In Section 4.3, we only preliminarily identify main-sequence binaries based on the HRD, and cannot entirely eliminate the influence of binaries. Consequently, the dispersion of $\Delta \, \feh$ for dwarfs is relatively larger than that for giants.

For this reason, we use the results of the giants as a reference. The dispersion of $\Delta\,\feh$ in our results is 0.065\,dex at $G = 12$, 0.18\,dex at $G = 16$, and 0.30\,dex at $G = 17$. However, these variances include the $\feh$ uncertainties from LAMOST, 
which vary by $SNR_g$, and are typically 0.05 dex for $SNR _g > 100$ (\citealt{Niu2021a}). Assuming that these uncertainties add in quadrature, a 0.065\,dex dispersion at $G = 12$ implies an uncertainty of 
0.04\,dex attributable to our result. 
In fact, the precision does not depend solely on brightness, but also on metallicity.  Thus, we test the precision of our results for different $\feh$ values using several star clusters in the next subsection. This approach also eliminates the effect of the $\feh$ uncertainties from LAMOST.

\subsection{Test by Clusters}
Our test is based on the member stars of four globular clusters (NGC 6397, NGC 6752, NGC 5904, and NGC 104) and two open clusters (NGC 2682 / M 67, NGC 6791). The four globular clusters are selected from those with the most member stars in the catalog provided by \cite{Baumgardt GC}, while the member stars of M 67 and NGC 6791 are taken from \cite{DR2 OC}. 
For all member stars, we only apply the criteria of $phot\_bp\_rp\_excess\_factor$ (fifth of the initial criteria) and $RUWE < 1.1$. 
For the five globular clusters, we only use giants for our test due to distance and brightness limitations. For M 67, we also include dwarfs, except for those identified as binaries in Section 4.3. 

\begin{figure*}[ht]
\includegraphics[width=180mm]{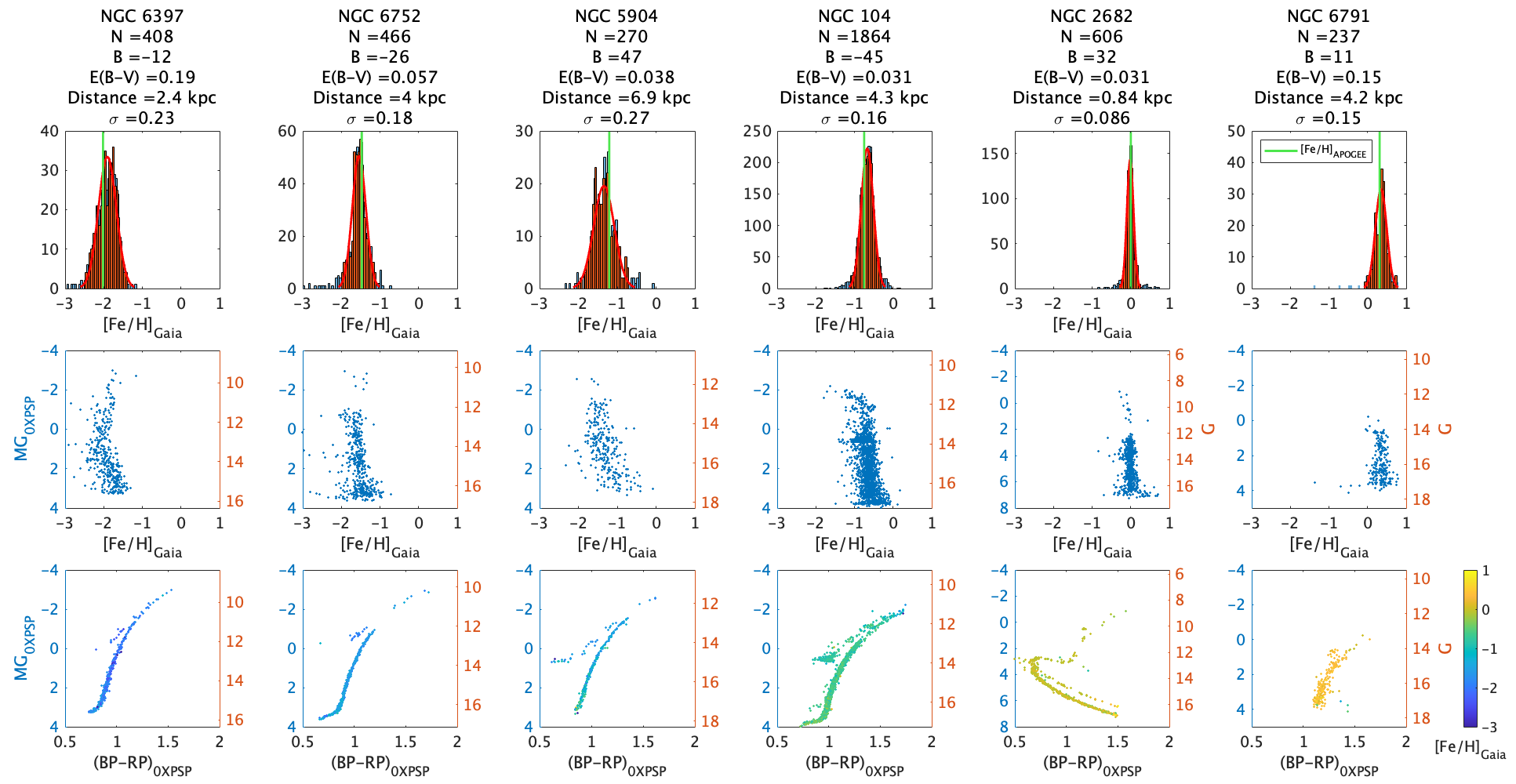}
\caption{Test results for star clusters. The title of each column sequentially lists the name of the cluster, the number of member stars, Galactic latitude, extinction, and distance. The first row of panels displays the metallicity distribution for member stars of each cluster.
Their standard deviations after multiple $3-\sigma$ clipping are labeled above each column. The median $\feh$ of the member stars in each cluster, as derived from APOGEE DR17, is indicated by green vertical lines.
The second row of panels depicts the distribution of $\feh_{\rm Gaia}$ from our results along with $MG_{0XPSP}$ and $G$, represented by the left and right vertical coordinates, respectively. 
The third row of panels shows the HRDs of these clusters, with the colors of the dots denoting $\feh_{\rm Gaia}$, as coded on the color bar on the right. The vertical coordinates are the same as those in the second row.} 
\label{Cluster}
\end{figure*}

The test results, shown in Figure~\ref{Cluster}, indicate that our photometric metallicity estimates exhibit good consistency for member stars of the clusters. 
The median $\feh$ values of the member stars in each cluster, as derived from our work, are $-$1.88, $-$1.54, $-$1.37, $-$0.66, $-$0.01, and +0.35, while the corresponding values from APOGEE DR17 are $-$2.02, $-$1.47, $-$1.20, $-$0.75, 0.00, and +0.31, respectively. Our results demonstrate good agreement with those provided by APOGEE.
Additionally, the panels in the second row of Figure~\ref{Cluster} show that the uncertainties increase as brightness decreases. 
Based on this test, for sources with $\feh \sim -2$, the precision of our result is about 0.25 dex when $G \sim 16$, while for sources with $\feh \sim 0$, the precision is about 0.09 dex under the same conditions.

\section{Properties of the Final catalog}

\subsection{The Final Metallicity Catalog}
We apply the models in Section 3 and the corrections in Section 4 to the 220 million Gaia XP sources to generate the final metallicity catalog. The columns contained in the final metallicity catalog are listed in Table \ref{columns}. 
Due to some sources falling outside the parameter range of our model, having poor observational quality, or suffering from inaccurate extinction estimates, our results may be unreliable for such stars. Therefore, we provide several flags to indicate the reliability of our results，such as the `Region’ and `Flag’ in the catalog. 
The `Region’ indicates that the source falls within the parameter range of our model and specifies the region from which the source’s photometric metallicity is derived. The `Flag’ denotes whether the result is reliable. For details, please refer to the caption of Table \ref{columns}.
It is worth mentioning that the `isLSMC' column in the catalog is derived based on the following criteria, involving positions and proper motions to classify whether a source comes from the Large or Small Magellanic Cloud. Since the distances of sources from the Magellanic Clouds are often incorrect, which can cause incorrect $MG_{0XPSP}$ values, we substitute all results from sources originating from the Magellanic Clouds with results from the Giant I region without the $MG_{0XPSP}$ model.

\begin{gather}
    LMC: \notag \\ 
    55^\circ<ra<105^\circ, -80^\circ<dec<-55^\circ  \notag \\
    0.3 <pmra < 3 \, (mas/yr) \notag \\
    -2 < pmdec < 2 \, (mas/yr)  \notag \\ 
    SMC: \notag \\ 
    0^\circ<ra<40^\circ, -80^\circ<dec<-65^\circ  \notag \\
    -0.5 <pmra < 2.5 \, (mas/yr) \notag \\ -2.5 <pmdec<-0.3 \, (mas/yr)  \notag \\ 
\end{gather}

\begin{figure}[ht]
\includegraphics[width=80mm]{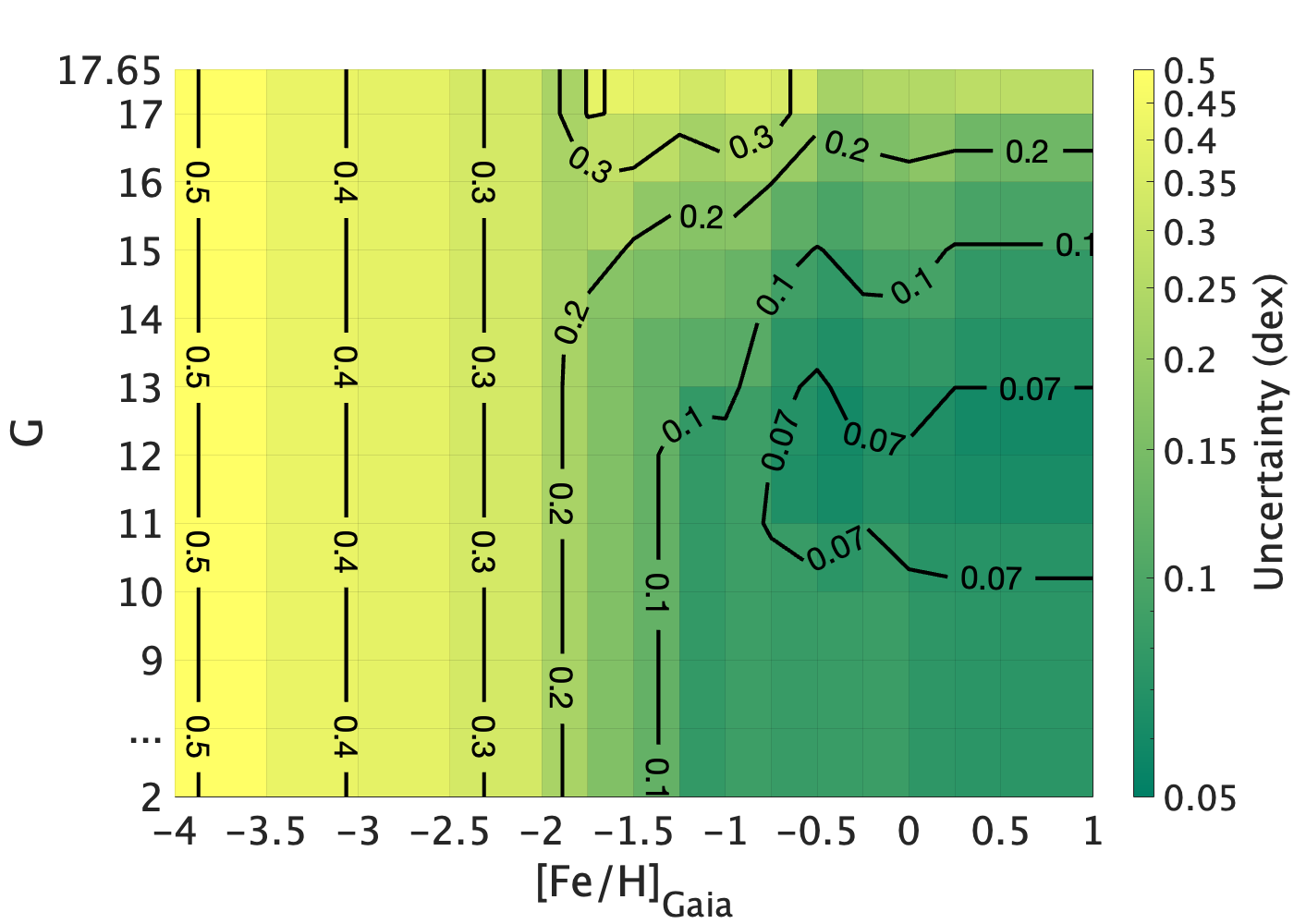}
\caption{The uncertainties of our results for different brightness and $\feh_{\rm Gaia}$ values, derived from the comparison with LAMOST and PASTEL. The black contours indicate the interpolated positions of the uncertainties, ranging from 0.07 to 0.5, as coded on the color bar on the right. } 
\label{precision}
\end{figure}

Based on the comparison with LAMOST DR7 in Section 5.3, we provide the typical uncertainties of our measurements for different brightness and $\feh_{\rm Gaia}$ values as references. 
Figure~\ref{precision} presents the uncertainties in our results, constrained by the brightness and $\feh$ ranges of the LAMOST sample (approximately $G > 10$ and $-1.75 < \feh < 0.5$). For brighter sources, we use the uncertainties at $G = 10$, and for $\feh_{\rm Gaia} > +0.5$, we refer to the uncertainties at $\feh_{\rm Gaia} = +0.5$. For sources with $\feh_{\rm Gaia} < -1.75$, we use the PASTEL sources in the reference sample as a reference. Due to the small number of sources, the brightness differences are not considered for sources with $\feh_{\rm Gaia} < -1.75$. The typical brightness of the PASTEL sources is $G \approx 11$ at $\feh_{\rm Gaia} = -2$ and $G \approx 13$ at $\feh_{\rm Gaia} = -3$. 
The values in the `Uncertainty' column of the final metallicity catalog are interpolated from the results shown in Figure~\ref{precision}.
The uncertainty is smallest when $G \approx 12$ and $\feh_{\rm Gaia} \approx 0$. It increases as the brightness and/or metallicity decreases. Note that, similar to the conclusion in section 5.3, this result incorporates the uncertainties in $\feh$ from LAMOST, which vary with $SNR_g$.

\setlength{\tabcolsep}{1.5mm}{
\begin{table*}[htbp]
\footnotesize
\centering
\caption{Description of the Final Catalog}
\begin{tabular}{lll}
\hline
\hline
Field & Description & Unit\\
\hline
source\_id & Unique source identifier for Gaia EDR3 (unique within a particular Data Release) & --\\
FEH & Photometric metallicity from this work  & --\\
FEH\_no\_HRD\_cor & Photometric metallicity from this work without HRD correction  & --\\
Region$^{a}$ & Flag to classify the result from different Regions & --\\
Flag$^{b}$ & Flag of reliability & --\\
rgeo & Median of the geometric distance posterior from \citet{BailerJonesDist} & pc\\
L & Galactic longitude from Gaia DR3 & deg\\
B & Galactic latitude from Gaia DR3 & deg\\
syn\_G\_cor & Synthetic photometry of G-band from corrected Gaia XP spectra & --\\
syn\_BP\_cor & Synthetic photometry of BP-band from corrected Gaia XP spectra & --\\
syn\_RP\_cor & Synthetic photometry of RP-band from corrected Gaia XP spectra & --\\
ra & Right ascension from Gaia DR3 & deg\\
dec & Declination from Gaia DR3 & deg\\
parallax & Parallax from Gaia EDR3 & mas\\
EBV & Value of \ebv$_{SFD}$ & --\\
pmra & Proper motion in right ascension direction from Gaia EDR3 & mas/year\\
pmdec & Proper motion in declination direction from Gaia EDR3 & mas/year\\
phot\_bp\_mean\_mag & Integrated BP-band mean magnitude from Gaia EDR3 & --\\
phot\_rp\_mean\_mag & Integrated RP-band mean magnitude from Gaia EDR3 & --\\
phot\_g\_mean\_mag & G-band mean magnitude from Gaia EDR3 & --\\
ruwe & Renormalised unit weight error from Gaia EDR3 & --\\
phot\_bp\_rp\_excess\_factor & BP/RP excess factor from Gaia EDR3 & --\\
isBinary & Flag to classify main-sequence binaries; 0 for not a binary and 1 for a binary & --\\
isLSMC & Flag to classify whether a source come from Large/Small Magellanic Cloud by positions and proper motions; & \\ 
   &  0 for No and 1 for Yes & --\\
Uncertainty & Uncertainties of \feh $\,\,$ estimated from the comparison with LAMOST and PASTEL & dex\\
\hline
\end{tabular}
\label{columns}
\tablecomments{(a)
The `Region' follows the format of ABC. A/B can be either 0 or 1. `A = 1' indicates that $MG_{0XPSP}$ is available, while `A = 0' not. 
`B = 1' signifies that the source falls within the parameter range of our model, whereas `B = 0' not. The value of C ranges from 1 to 5, denoting the specific region from which the source's photometric metallicity is derived: `C = 1' for the Dwarf Region, `C = 2' for the Giant I Region without $MG_{0XPSP}$, `C = 3' for the Giant I Region with $MG_{0XPSP}$, `C = 4' for the RC-HB Region, and `C = 5' for the Giant II Region. A simple filtering condition, `Region $>$ 110', allows for the selection of samples that fall within the model parameters range.
}
\tablecomments{(b)
The `Flag' follows the format of ABCDE, where each letter can be either 0 or 1. A, B, C, D, and  E, respectively indicates whether the following criteria are met: $phot\_bp\_rp\_excess\_factor <0.02 \times (BP_{XPSP}-RP_{XPSP})^2 + 0.055 \times (BP_{XPSP}-RP_{XPSP}) + 1.165$, $RUWE < 1.1$, \ebv $< 0.5$, absolute values of Galactic latitude greater than 20\degr, and vertical distances to the Galactic disk greater than 0.2 kpc. For example, `Flag = 11111' means that the source meets all of the above criteria.}

\end{table*}}

\vskip 1cm
\subsection{Statistical Properties}

Based on the final metallicity catalog, we analyze the statistical properties of our results after applying a simple filtering using the columns `Region' and `Flag'. 
Out of a total of 220 million sources, approximately 175 million sources have $MG_{0XPSP}$ available and fall within the parameter ranges of the models (`Region $>$ 110'). When $MG_{0XPSP}$ is not available, sources with $0.6 < (BP-RP){0XPSP} < 1.45$ use the Giant I model without $MG{0XPSP}$, while the other sources use the Dwarf model.
Further, we consider all criteria except for low Galactic latitude (`Flag = `11101', `isBinary = 0', and `isLSMC = 0') to ensure photometric quality, correct extinction, and to remove likely metal-rich binaries. 
Approximately 82 million sources remain for the analyses in this subsection. It should be noted that $RUWE <1.1$ represents a stringent criterion, resulting in the exclusion of nearly 30 million sources. Thus, this criterion may need adjustment based on the specific science case.

\begin{figure*}[ht]
\includegraphics[width=180mm]{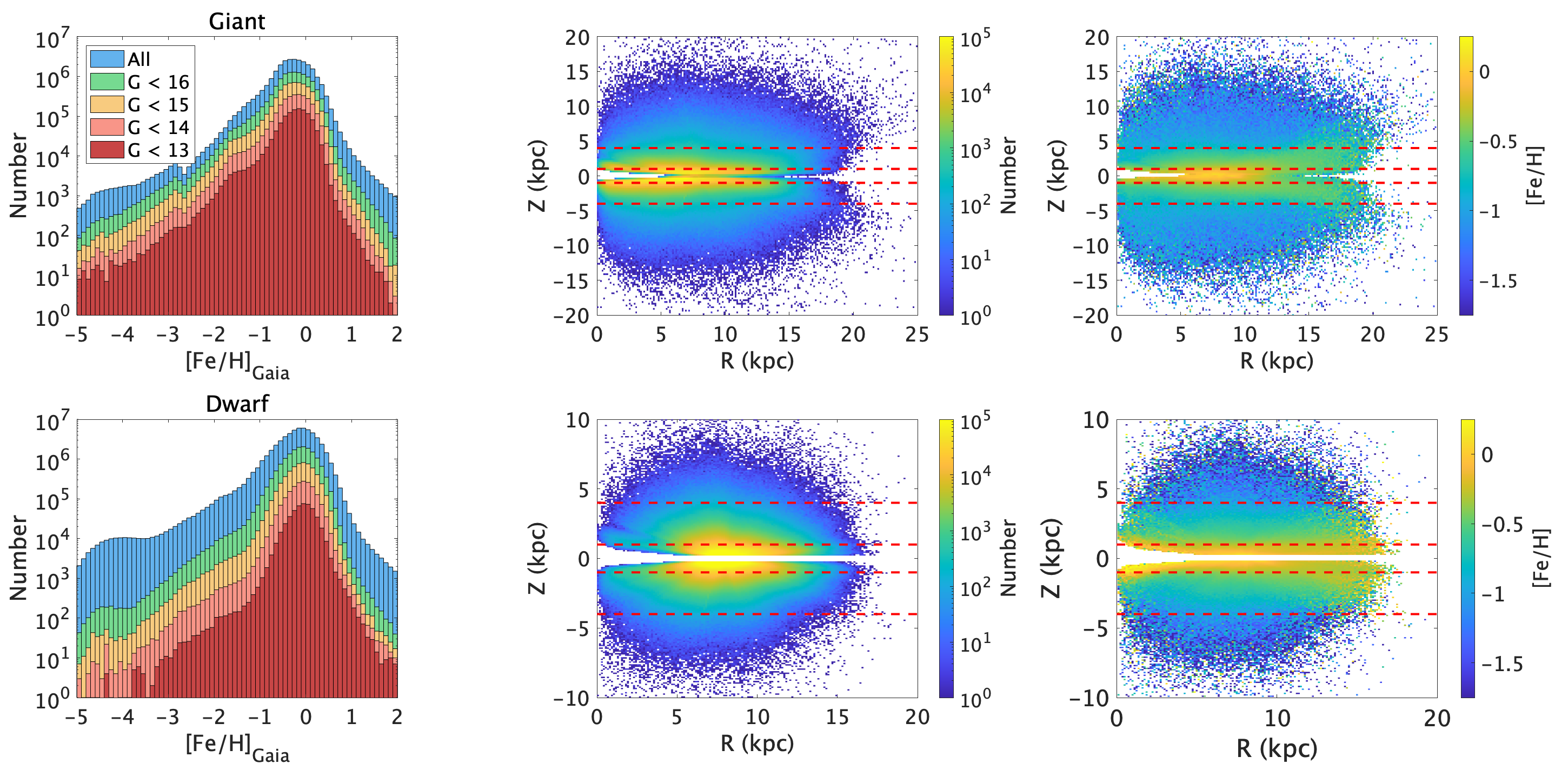}
\caption{Statistical properties of the stars selected for the final catalog. The two left panels show the $\feh_{\rm Gaia}$ distribution for giants and dwarfs, respectively. The middle and right panels depict the spatial distributions in the R $-$ Z plane，color-coded by number densities and $\feh_{\rm Gaia}$, respectively. The Sun is located at (R, Z) = (8.178, 0.0) kpc. The red horizontal dashed lines mark $Z = \pm 1$\,kpc and $Z = \pm 4$\,kpc.
}
\label{Sta_1}
\end{figure*}

The $\feh_{\rm Gaia}$ distribution of this trimmed sample is displayed in the two left panels of Figure~\ref{Sta_1}. The discontinuity in the giant sub-sample at $\feh_{\rm Gaia} = -2.75$ arises from the limitation of the range of the correction on the HRD. This correction reduces the contamination more effectively for $\feh_{\rm Gaia} > -2.75$ than for $\feh_{\rm Gaia} < -2.75$. As the brightness increases, the contamination at the metal-poor end decreases, thus the discontinuity gradually weakens and eventually disappears.
For dwarfs, the number of VMP stars is increased due to likely contamination by binaries. Therefore, careful selections are required in studies involving VMP stars.
Based on this sample, our catalog includes 705,937 VMP stars and 224,028 EMP stars. For $G < 16$, it includes 93,918 VMP stars and 18,139 EMP stars; for $G < 15$, it includes 37,524 VMP stars and 6,501 EMP stars; and for $G < 14$, it includes 13,284 VMP stars and 2,132 EMP stars.

The spatial distributions in the R $-$ Z plane are shown in the middle and right panels of Figure~\ref{Sta_1}, where R denotes the Galactocentric distance in cylindrical coordinates, and Z represents the distance above or below the Galactic plane. 
The middle and right panels are color-coded by number densities and $\feh_{\rm Gaia}$ （median value of each bin, respectively. 
For both giants and dwarfs, clear structures that may be associated with the thin and thick disks are evident in the right panels. Specifically, clear boundaries can be seen between the thin disk and the thick disk at $Z \approx 1\,kpc$, and between the thick disk and the halo at $Z \approx 4\,kpc$. Additionally, at $R \approx 12$  kpcand beyond, a noticeable flare of the Galactic disk is observed. 
For halo stars, we have confirmed that the metallicity progressively decreases with Galactocentric distance in spherical coordinates from 5 to 30 kpc, with a slope of approximately $-$0.0256 dex/kpc. The typical $\feh_{\rm Gaia}$ values are $-$0.916 at 5 kpc and $-$1.300 at 20 kpc.

\begin{figure*}[ht]
\includegraphics[width=180mm]{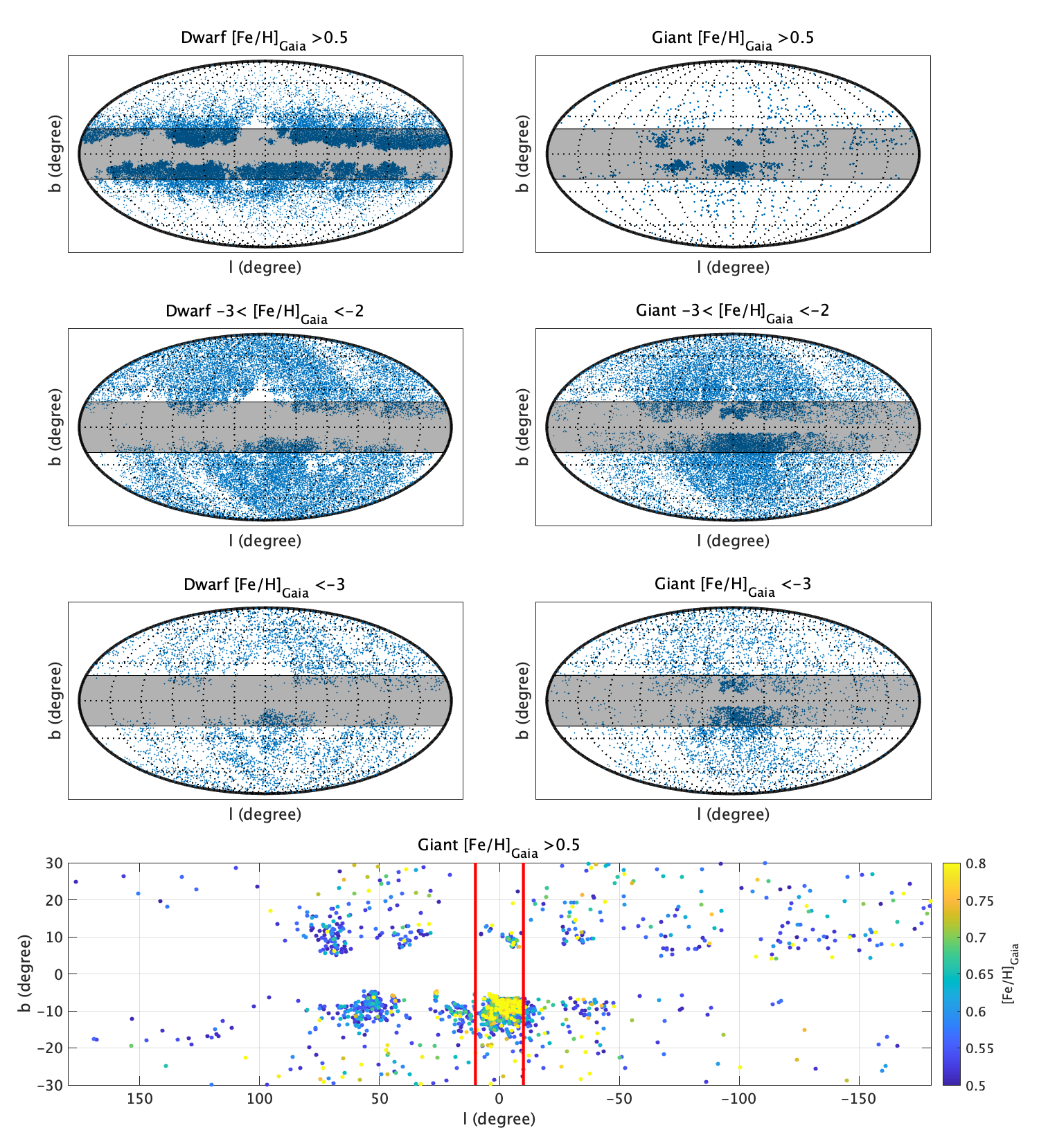}
\caption{The spatial distributions of extremely metal-rich ($\feh_{\rm Gaia} > +0.5$) and metal-poor stars ($-3<\feh_{\rm Gaia}<-2$ and $\feh_{\rm Gaia} <-3$), shown in Galactic coordinates centered on the Galactic center. Shaded areas indicate regions where the Galactic latitude is lower than 20$^\circ$. 
Foreground dwarfs closer than 1 kpc and giants closer than 4 kpc from the Sun are removed.
The bottom panel is same as the panel in the upper right corner, but zoomed in and color-coded by $\feh_{\rm Gaia}$, as shown in the color bar on the right. The two vertical lines represent galactic longitudes of 10$^\circ$ and -10$^\circ$.} 
\label{Sta_2}
\end{figure*}

We further investigate the spatial distributions for extremely metal-rich (EMR; $\feh_{\rm Gaia} > +0.5$), VMP, and EMP stars in Figure~\ref{Sta_2}.
To minimize contamination, we exclude faint sources with $G > 16$.
For EMR stars, we additionally require $(BP-RP)_{0XPSP} < 1.4$ to eliminate sources from the low sensitivity range of stellar loci， and $\ebv < 0.3$ to ensure the accuracy of the SFD98 map.
For VMP and EMP dwarfs, we only select main-sequence turnoff stars ($0.5 <(BP-RP)_{0XPSP} <0.8$). 
Two very recent works (\citealt{EMRknot, EMRXP}) report a central `knot' of EMR stars. To explore this structure, we follow \cite{EMRXP} to remove foreground giants closer than 4 kpc from the Sun and foreground dwarfs closer than 1 kpc from the Sun. 
For VMP and EMP giants, we select stars with $0.7 <(BP-RP)_{0XPSP} <1.8$ and exclude stars significantly off the isochrones (the solid green curves in Figures \ref{Giant_overall_HRD_sct} and \ref{Giant_overall_HRD_grd}. 

The results for the spatial distributions of EMR, VMP, and EMP stars are shown in Figure~\ref{Sta_2}. 
The first row displays the result of EMR stars.  Notably,  we observe a similar `knot' structure concentrated in the direction of the Galactic center for giants, consistent with \cite{EMRXP}. As shown in the bottom panel of the Figure~\ref{Sta_2}, most metal-rich sources of our result are concentrated quite sharply within the central 10$^\circ$ (marked by the two vertical red solid lines), which is also consistent with \cite{EMRXP}.
The north-south asymmetry is due to the removal of regions with high extinction. 
On the other hand, metal-poor stars (shown in the second and third rows of Figure~\ref{Sta_2}) are predominantly distributed in the halo. For giants, the concentration is also visible in the direction of the Galactic center, consistent with the metal-poor `heart' of the Galaxy （\citealt{PoorOldHeart}）. 

\section{Summary}
In this work, we have generated a photometric metallicity catalog for about 100 million stars based on the Gaia DR3 XP spectra，including 34 million giants with $0.6 < (BP-RP)_0 < 1.75$ and 65 million dwarfs with $0.2 < (BP-RP)_0 < 1.5$ ($|Z| >200\,pc$ and $\ebv <0.5$).  The catalog is publicly available at \href{url}{https://doi.org/10.12149/101548}.
These estimates are derived using the stellar loci fitting technique, primarily based on three broad-band synthetic photometric magnitudes: $G$, $BP$, and $RP$, and supplemented by Gaia DR3 distances. We have achieved precision comparable to that of the full Gaia XP spectra-based results. 
The typical precision is between 0.04 -- 0.1\,dex for both dwarfs and giants at [Fe/H] = 0 as faint as G $\sim$ 16, and decreases to 0.15 -- 0.25\,dex at [Fe/H] = $-$2.0, as validated by a number of tests, including comparison with members of star clusters. 
By restricting the relationship between the polynomial coefficients, the polynomial model maintains good robustness and discrimination at $\feh < -2.5$ and down to $\feh \sim -4$, as demonstrated by several tests, which significantly improves the success rate for selecting VMP and EMP stars (Xu et al., in preparation). Our catalog includes 93,922 VMP stars and 17,282 EMP stars with $G < 16$,  37,056 VMP stars and 6,192 EMP stars with $G < 15$, and 13,337 VMP stars and 2,030 EMP stars with $G < 14$. We point out that our results are influenced by carbon enhancement. We expect that by comparison with results derived by techniques that are not affected by carbon enhancement, we can effectively identify carbon-enhanced stars.

This work also demonstrates the feasibility and great potential of high-precision photometry for metallicity estimation. A number of lessons have been learned, including:
\begin{enumerate}
    \item  
    Low sensitivity optical broad-band photometry can also estimate metallicity precisely, provided that the photometric precision is sufficiently high. The precision of the photometirc metallicity (and presumably other stellar-atmospheric parameters) estimation is the combined  result of sensitivity and photometric uncertainty (\citealt{XiaoFilter}).
    \item Broad-band photometry is influenced by carbon abundance when the filter transmission curves cover the strong carbon absorption features.
    \item Different types of stars (e.g., red clump stars and red giant stars) have distinct stellar loci, and the discrepancies can no longer be neglected at the current precision.
    \item Absolute magnitude can provide metallicity information in regions where the stellar loci have little to no sensitivity.
    \item Polynomial models with reasonable coefficient constraints can maintain robustness and discrimination in the VMP and EMP regions.
\end{enumerate}

\vfill\eject
\appendix
\setcounter{secnumdepth}{0}
\setcounter{table}{0}   
\setcounter{figure}{0}
\renewcommand{\thetable}{A\arabic{table}}
\renewcommand{\thefigure}{A\arabic{figure}}

\section{A. The $f_2$ Term for the Dwarf Region}
The $f_2$ term for the Dwarf region is presented in Table \ref{Table2}.

\begin{table}
\caption{The $f_2$ Term for the Dwarf Region}
\label{Table2} 
\centering
\begin{tabular}{*{7}{r}} \hline\hline
$Color1$ & $Color2$ & $Color1$ & $Color2$ & $Color1$ & $Color2$& \\ \hline
0.200 &0.06885 &0.650 &0.24774 &1.100 &0.47080 & \\
0.225 &0.07769 &0.675 &0.25905 &1.125 &0.48425 & \\
0.250 &0.08666 &0.700 &0.27053 &1.150 &0.49779 & \\
0.275 &0.09575 &0.725 &0.28214 &1.175 &0.51143 & \\
0.300 &0.10496 &0.750 &0.29393 &1.200 &0.52516 & \\
0.325 &0.11430 &0.775 &0.30585 &1.225 &0.53898 & \\
0.350 &0.12376 &0.800 &0.31791 &1.250 &0.55290 & \\
0.375 &0.13335 &0.825 &0.33015 &1.275 &0.56691 & \\
0.400 &0.14307 &0.850 &0.34245 &1.300 &0.58101 & \\
0.425 &0.15292 &0.875 &0.35489 &1.325 &0.59519 & \\
0.450 &0.16289 &0.900 &0.36738 &1.350 &0.60945 & \\
0.475 &0.17299 &0.925 &0.37998 &1.375 &0.62380 & \\
0.500 &0.18322 &0.950 &0.39267 &1.400 &0.63823 & \\
0.525 &0.19359 &0.975 &0.40542 &1.425 &0.65275 & \\
0.550 &0.20412 &1.000 &0.41825 &1.450 &0.66736 & \\
0.575 &0.21481 &1.025 &0.43119 &1.475 &0.68205 & \\
0.600 &0.22565 &1.050 &0.44426 &1.500 &0.69682 & \\
0.625 &0.23662 &1.075 &0.45747 & \\
\hline
\end{tabular}
$Color1$ and $Color2$ represent $(BP-RP)_{0XPSP}$ and $(BP-G)_{0XPSP}$, respectively.
\end{table}

\section{B. Comparisons with Metallicity Derived from High-Resolution Spectra}
\begin{figure*}[ht]
\includegraphics[width=180mm]{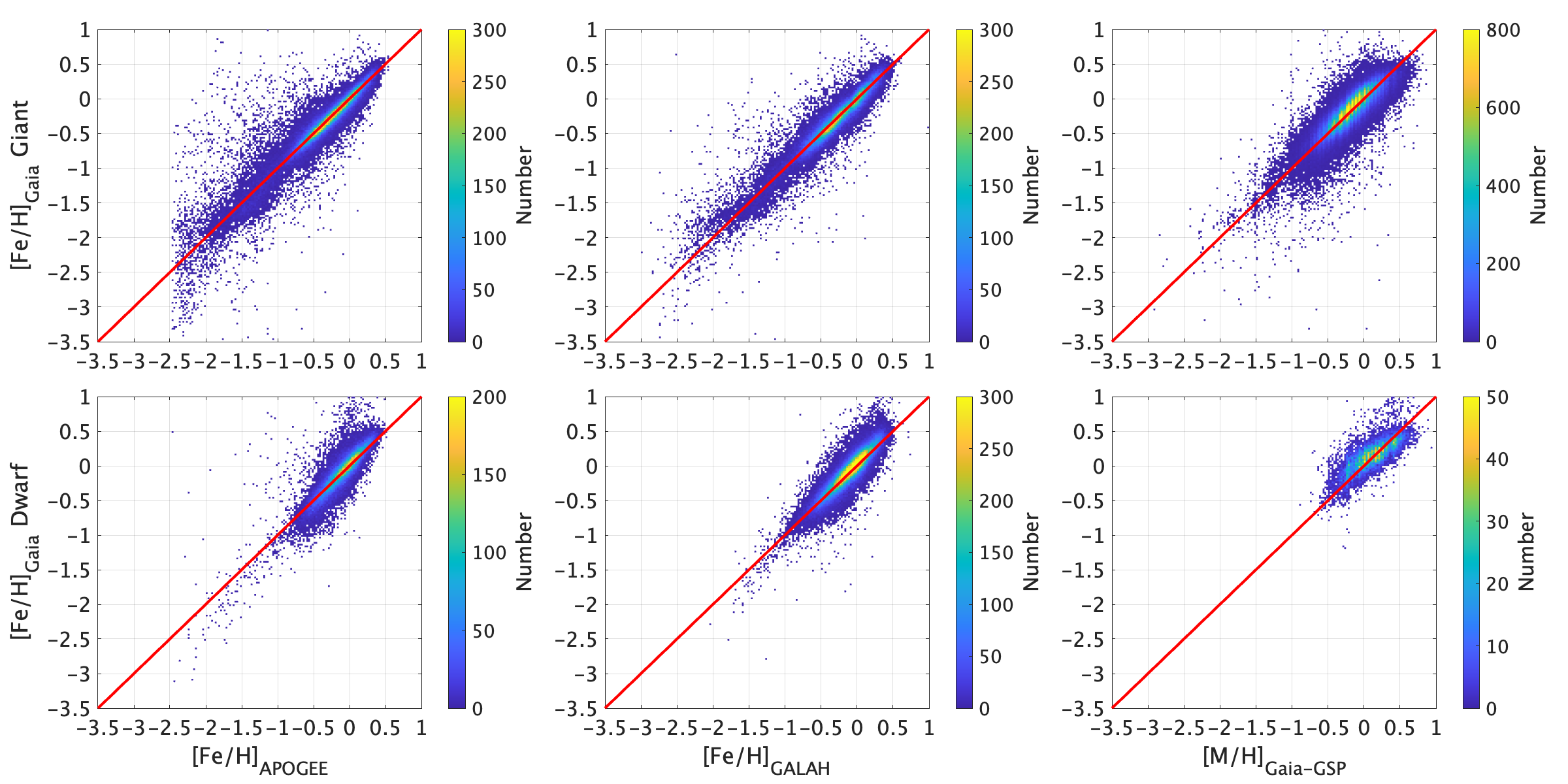}
\caption{Comparison of our metallicity estimates with those from APOGEE DR17, GALAH DR4, and Gaia DR3 GSP-spec. The top and bottom panels are for giants and dwarfs, respectively. The colors represent the density of points.}
\label{ADD1}
\end{figure*}

To further test and verify the accuracy and precision of our results, we compare our metallicity estimates with those derived from high-resolution spectra, including APOGEE DR17 (\citealt{APOGEEDR17}), GALAH DR4 (\citealt{GALAHDR4}), and Gaia DR3 GSP-Spec (\citealt{GaiaGSP2023}). 
For our results, we apply the same criteria as described in Section 5.1 to exclude sources with unreliable results. For the metallicity estimates derived from high-resolution spectra, we use different criteria to ensure the reliability of the selected sources. Specifically, for the APOGEE sample, we require signal-to-noise ratios greater than 50. For the GALAH sample, we use their recommended flag values: \texttt{snr\_px\_ccd3 > 30}, \texttt{flag\_sp = 0}, and \texttt{flag\_fe\_h = 0}. For the Gaia GSP-Spec sample, we require \texttt{vbroadM = 0}, \texttt{vradM = 0}, and \texttt{fluxNoise = 0}.
Based on the aforementioned selection criteria, there are 113,623, 155,329, and 234,734 common sources between our results and those from APOGEE, GALAH, and Gaia GSP-Spec, respectively. The comparison results are shown in Figure~\ref{ADD1}, demonstrating that our metallicity estimates are in good agreement with those derived from the aforementioned high-resolution spectra.
We further compare the dispersions of the differences between our metallicity estimates and those from high-resolution spectra for sources of $\feh > -0.5$. For giants, the dispersions relative to APOGEE, GALAH, and Gaia GSP-spec are 0.061 dex, 0.079 dex, and 0.10 dex, respectively. For dwarfs, the dispersions are 0.088 dex, 0.090 dex, and 0.14 dex, respectively.

\section{C. The influence of extinction}
\begin{figure*}[ht]
\includegraphics[width=180mm]{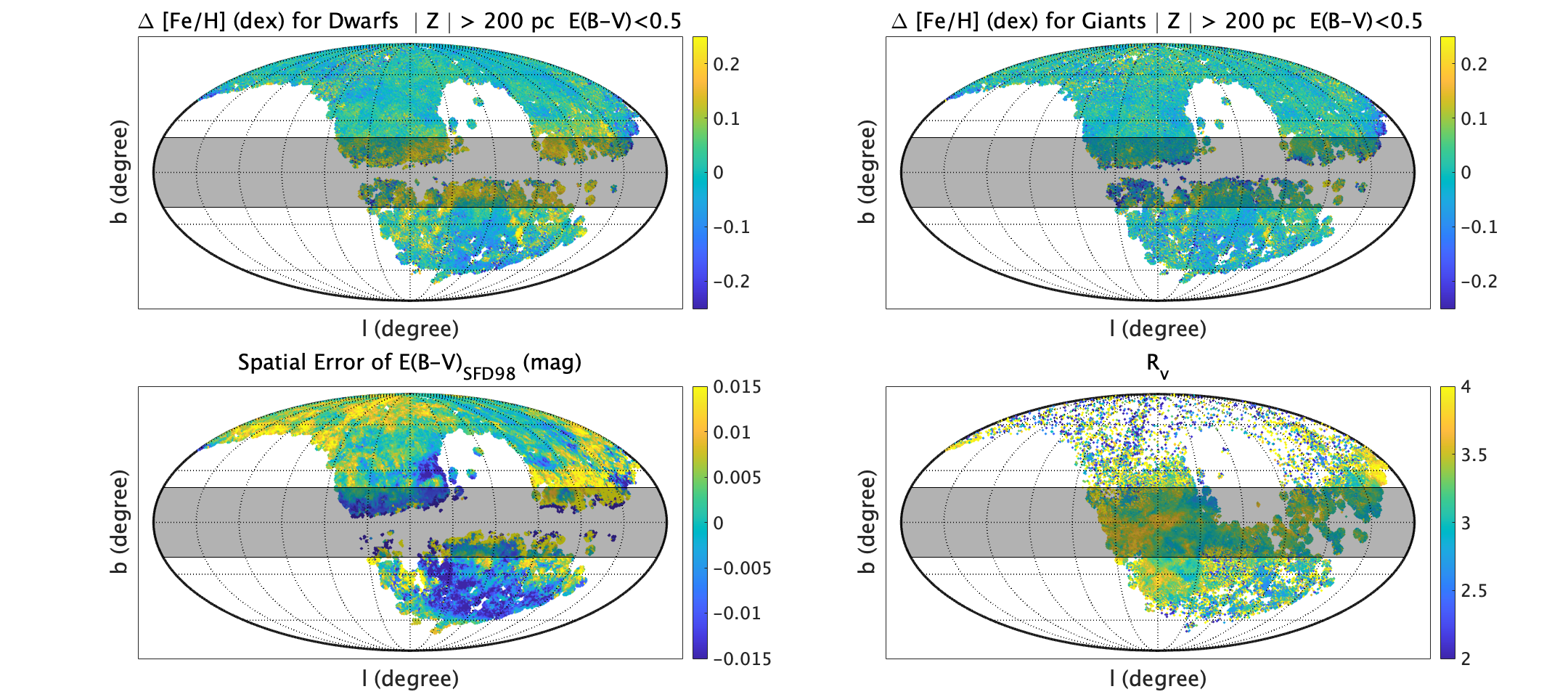}
\caption{The spatial distribution of discrepancies in metallicity estimates between our results and LAMOST DR7 (upper panels), along with the spatial errors of SFD98 and $R_\text{V}$, shown in Galactic coordinates centered on the Galactic anti-center (lower panels). Shaded areas indicate regions where the Galactic latitude is lower than 20$^{\circ}$. The colors represent the number density of points.} 
\label{Spatial_1}
\end{figure*}

\begin{figure}[ht]
\includegraphics[width=90mm]{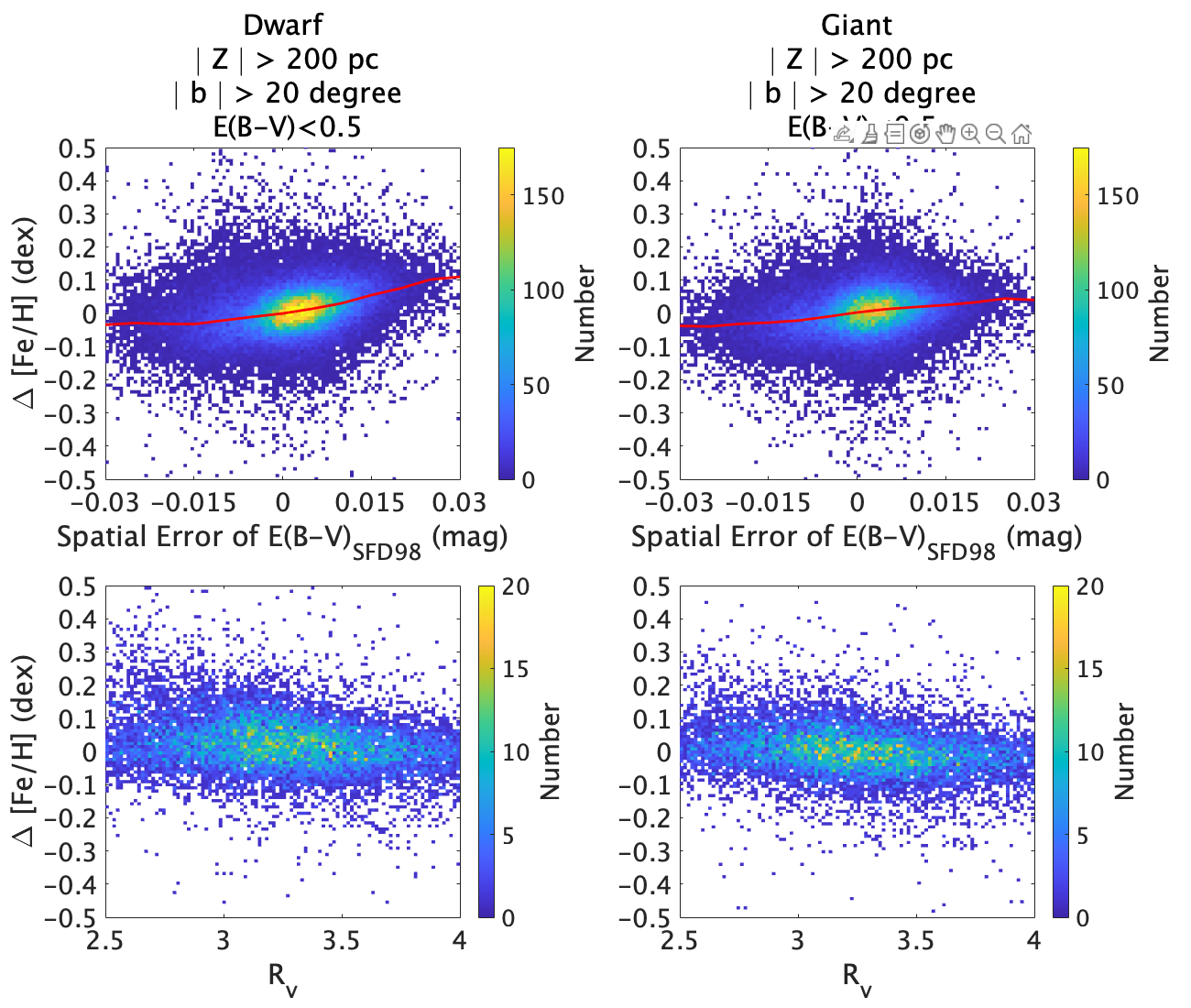}
\caption{The correlation between the spatial errors of SFD98 and $\Delta \feh$, as well as the correlation between $R_\text{V}$ and $\Delta \feh$. 
The red solid lines in the top two panels indicate the median $\Delta \feh$ corresponding to different values for the errors of \ebv. The colors represent the number density of points.} 
\label{Spatial_2}
\end{figure}

\begin{figure*}[ht]
\includegraphics[width=180mm]{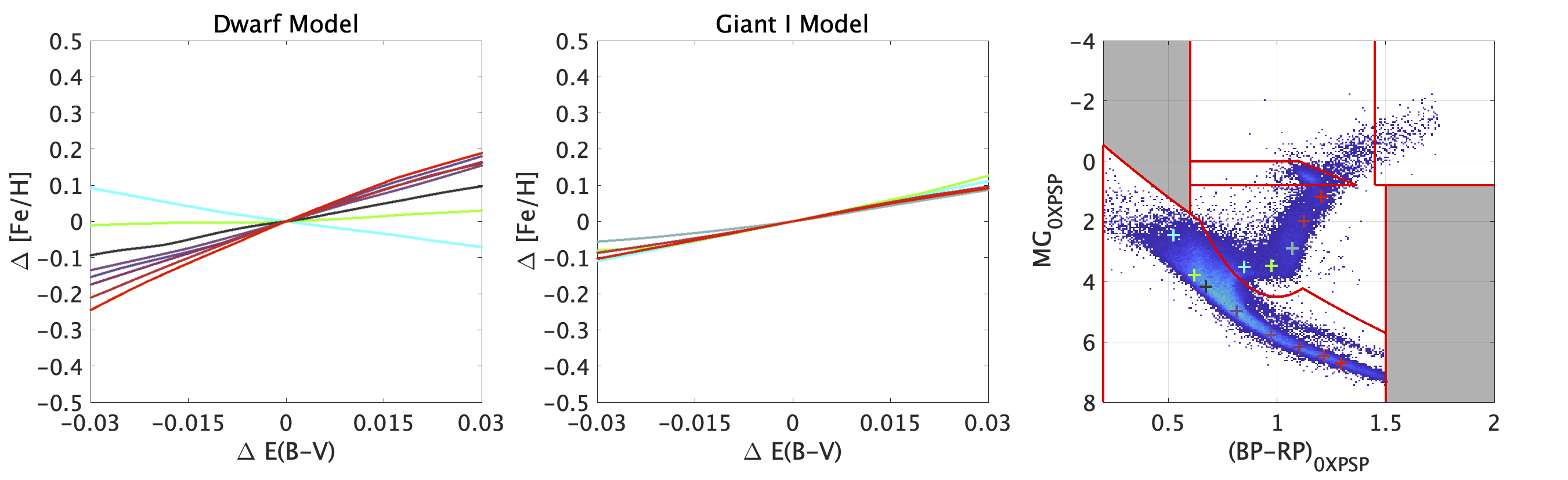}
\caption{The correlation between $\Delta\,$\ebv\ and $\Delta\, \feh$ derived by selected sources with varying \ebv. The rightmost panel displays the HRD of the sample, with cross markers indicating the selected sources. The two panels on the left depict the correlation for the selected sources in the Dwarf and Giant models, respectively. The color of the lines represents different $(BP-RP)_{0XPSP}$ of the sources, which corresponds to the color of the cross markers in the rightmost panel.} 
\label{Spatial_3}
\end{figure*}

In Section 4.1, we mentioned that accurate extinction corrections are essential for deriving correct photometric metallicity estimates. The SFD98 dust reddening map we used exhibits moderate spatially dependent errors (\citealt{SunSFD}). Additionally, the extinction coefficients from \cite{ZhangRYExtinction2023} are the typical values at $R_\text{V} \sim 3.1$, but $R_\text{V}$ is also known to vary spatially (\citealt{ZhangRVmap}). Therefore, we explore the impact of the spatial errors in SFD98 and the spatial variations of $R_\text{V}$ on our results.

As in section 5.3, we compare our result with LAMOST DR7, and subsequently partition the celestial sphere into 196,608 tiles by applying the HEALPix (\citealt{HealpixPaper,HealpixTool}) scheme (nside = 128, corresponding to a spatial resolution of $\approx$ 27.'5). The median discrepancies between our result and LAMOST DR7 （$\Delta \feh$） are shown in Figure~\ref{Spatial_1}. We also include the spatial errors of SFD98 from \cite{SunSFD} and the $R_\text{V}$ map from \cite{ZhangRVmap} for comparison.
Figure~\ref{Spatial_1} shows that $\Delta\, \feh$ exhibits a comparable structure with spatial errors of SFD98.
Thus, we directly examine the correlations between $\Delta\, \feh$ and the spatial errors of SFD98 for dwarfs and giants in the top panels of Figure~\ref{Spatial_2}. The correlation is more pronounced for the dwarfs, consistent with Figure~\ref{Spatial_1}. We further examine the correlations between $\Delta\, \feh$ and $R_\text{V}$ after correcting for the correlation with the spatial errors of SFD98 (represented by the red solid lines). The results indicate a weak correlation, but the overall effect is minimal.

To directly test our models, a sub-sample of relatively metal-rich ($\feh_{\rm LAMOST} > -0.75$) stars with low extinction values between $0.03 < ~\ebv~ < 0.06$ is selected. 
Within this sub-sample, we select stars with different colors, apply varying \ebv~ values, and then compare the correlation between $\Delta\,$\ebv~ and $\Delta\, \feh$~ in Figure~\ref{Spatial_3}. 
It is evident that there is a correlation between $\Delta\,$\ebv~ and $\Delta\, \feh$ for both Dwarf and Giant models, and this correlation depends on color.  Furthermore, the correlations shown in Figure~\ref{Spatial_3} are broadly consistent with the results in Figure~\ref{Spatial_2}, suggesting that the discrepancies between our result and LAMOST DR7 are due to spatial errors in extinction.

\vspace{7mm} \noindent {\bf Acknowledgments}
H.Y. dedicates this paper to his grandmother, Chunyan He (1930 -- 2025). 
The authors thank the anonymous referee for his/her suggestions that improved the clarity of our presentation. This work is supported by the National Key R\&D Program of China via 2024YFA1611901 and 2024YFA1611601, and the National Natural Science Foundation of China through the projects NSFC 124B2055, 12222301, and 12173007. M.X. acknowledges financial support from the National Key R\&D Program of China through grant no. 2022YFF0504200.
T.C.B. acknowledges partial support from Physics Frontier Center/JINA Center for the Evolution of the Elements (JINA-CEE), and OISE-1927130: The International Research Network for Nuclear Astrophysics (IReNA), awarded by the US National Science Foundation (NSF).

This work has made use of data from the European Space Agency (ESA) mission {\it Gaia} (\url{https://www.cosmos.esa.int/gaia}), processed by the Gaia Data Processing and Analysis Consortium (DPAC, \url{https:// www.cosmos.esa.int/web/gaia/dpac/ consortium}). Funding for the DPAC has been provided by national institutions, in particular the institutions participating in the Gaia Multilateral Agreement. 

The Guoshoujing Telescope (the Large Sky Area Multi-Object Fiber Spectroscopic Telescope LAMOST) is a National Major Scientific Project built by the Chinese Academy of Sciences. Funding for the project has been provided by the National Development and Reform Commission. LAMOST is operated and managed by the National Astronomical Observatories, Chinese Academy of Sciences.

{}

\end{CJK*}
\end{document}